\documentclass{iopart}

\usepackage{iopams}  
\usepackage{setstack}  
\usepackage{graphicx}
\usepackage{bbm,bm}

\graphicspath{{eps/},{~/investigacion/tesis/eps}}

\def\>{\rangle} \def\<{\langle} 

\def\e{{\rm e}}
\def\H{{\rm H}}
\newcommand{\ie}{\textit{i.e.} }
\newcommand{\eg}{\textit{e.g.} }
\newcommand{\Eg}{\textit{E.g.} }

\newcommand{\mimath}{\rmi}

\newcommand{\intoh}{\int_0^t \rmd \tau \int_0^\tau \rmd \tau'}

\providecommand{\openone}{\leavevmode\hbox{\small1\kern-3.8pt\normalsize1}}

\newcommand{\tre}{\tr_{\rm e}}
\newcommand{\mcH}{\mathcal{H}}
\newcommand{\mcO}{\mathcal{O}}
\newcommand{\mcU}{\mathcal{U}}

\begin{document}
\title[Decoherence of two qubit systems: A random matrix description]{Decoherence of two qubit systems:\\ A random matrix description}
\author{C Pineda$^{1,2,3}$, T Gorin$^{3,4}$, and T H Seligman$^{1,3}$}
\address{$^1$Instituto de Ciencias F\'{\i}sicas, Universidad Nacional Aut\'onoma de M\'exico, M\'exico}
\address{$^2$Instituto de F\'{\i}sica, Universidad Nacional Aut\'onoma de M\'exico, M\'exico}
\address{$^3$Centro Internacional de Ciencias, Cuernavaca, M\'exico}
\address{$^4$Departamento de F{\'\i}sica, Universidad de Guadalajara S. R. 500, 44420 Guadalajara, Jalisco, M\'exico}
\ead{carlospgmat03@gmail.com}

\begin{abstract}
We study decoherence of two non-interacting qubits.  The environment and its
interaction with the qubits are modelled by random matrices. Decoherence,
measured in terms of purity, is calculated in linear response approximation.
Monte Carlo simulations illustrate the validity of this approximation and of
its extension by exponentiation.  The results up to this point are also used to
study one qubit decoherence.  Purity decay of entangled and product states are
qualitatively similar though for the latter case it is slower.  Numerical
studies for a Bell pair as initial state reveal a one to one correspondence
between its decoherence and its internal entanglement decay.  For strong and
intermediate coupling to the environment this correspondence agrees with the
one for Werner states.  In the limit of a large environment the evolution
induces a unital channel in the two qubits, providing a partial explanation for
the relation above.
\end{abstract}

\section{Introduction}

In recent experiments \cite{NatureTomography, fourparticleentanglement,
photonentanglement, antidecoherenceentanglement} it has been demonstrated that
it is possible to protect ever larger entangled quantum systems, often arrays
of qubits, ever more efficiently from decoherence.  This improved protection of
states makes a more detailed analysis of decoherence desirable. In the context
of fidelity decay, a random matrix description \cite{1367-2630-6-1-020,
reflosch} is accessible and very effective in describing experiments
\cite{expRudi, 1367-2630-7-1-152, gorin-weaver}. A close connection between the
dynamics of fidelity decay and decoherence has been shown in some instances
\cite{Zur91,gpss2004}, which suggests to apply methods successful in one field
to the other. Based on some previous work we shall therefore analyse in detail
the consequences of using random matrix theory (RMT) to model decoherence
\cite{reflosch, 1464-4266-4-4-325, pinedaRMTshort} for the case of two
non-interacting qubits or a single qubit as the central system.

Two perspectives make such a random matrix treatment particularly attractive.
First, reduction of decoherence may, in some instances, be achieved by
isolating some ``far'' environment (including spontaneous decay) to a degree
that it can, to first approximation, be neglected. Then it can happen that the
Heisenberg time of the relevant ``near'' environment is finite on the time
scale of decoherence.  In such a case it becomes relevant that RMT shows, in
linear response approximation, a transition from linear to quadratic decay at
times of the order of the Heisenberg time. This behaviour is seen with spin
chain environments \cite{privatepineda2006}, and is essential for the success
of the theory in describing the above mentioned experiments of fidelity decay.
Note also that the concept of a two stage environment has been used for basic
considerations \cite{zurekreview}.  Second, the long term goal must be to
describe in one theory the decay of fidelity that includes undesirable
deviations of the internal Hamiltonian of the central system, 
together with decoherence.

This study is focussed on weak coupling of one or two
qubits to an environment, and thus we can use the correlation function approach
proposed for purity decay in echo-dynamics \cite{purityfidelity}, treating
the coupling as the perturbation \cite{reflosch}. The linear response approximation
will be sufficient and in this approximation the ensemble averages, which we
have to take in any RMT model, are feasible though somewhat tedious. 
Exact solutions, which exist in some
instances for the decay of the fidelity amplitude \cite{shortRMT,gorin:244105},
seem to be out of reach at present, because they would require the evaluation
of four-point functions.

Assume that the qubit(s) are initially in a pure state, and evolve under their
own local Hamiltonians.  The qubit(s) are coupled independently via random
matrices to a large environment in turn described by another random matrix. The
coupling to the environment gives rise to decoherence.  Averaging both the
coupling and the environment Hamiltonian over the RMT ensembles yields the
generic behaviour.  We then focus on the correspondence between decoherence and
entanglement decay for a Bell state.  We find a relation between this two
quantities, that can be partially understood assuming unitality.  This relation
in most cases coincides with the corresponding one obtained for Werner states.
For that part, it is essential that the two qubits do not interact.  Otherwise,
the coupling between the qubits would act as an additional sink (or source) for
internal entanglement -- a complication we wish to avoid.  However, a residual
coupling between the qubits should be taken into account at some
point~\cite{Pineda01}.  A preliminary study of some aspects of this system has
been presented in~\cite{pinedaRMTshort}, but here we shall present a more
general picture: First, we allow for local one-qubit Hamiltonians. They have a
considerable effect on decoherence.  Second, we cover exhaustively {\it all}
possible initial states. Third, we study the consequences of the whole system
being time-reversal invariant.  Indeed, we present the mathematical procedure
in detail using both the Gaussian unitary (GUE) and the Gaussian orthogonal
(GOE) ensembles~\cite{cartanRMT,mehta} for the description of the environment and the coupling.  The
two ensembles correspond to time reversal invariance (TRI) breaking and
conserving dynamics respectively. All these cases are treated on equal footing.

In \sref{sec2} we shall state the model,  recall the linear response formalism
for echo dynamics, and show how it can be adapted to forward evolution. In
\sref{O} we shall discuss decoherence of a single qubit, which already shows
some interesting features by itself and serves as a warm up for the more
complicated case of two qubits.  The details of the calculations are given in
the appendices. This shall be treated in \sref{O2}, starting from the situation
where one qubit is a spectator, exerting influence only through entanglement.
More general cases are analysed and reduced to the spectator situation.  Most
analytic results will be accompanied by Monte Carlo simulations.   In \sref{R}
we investigate the correspondence between decoherence (purity) and entanglement
(concurrence) as both evolve in time. We end with conclusions in
\sref{sec:conclusions}.

\section{General considerations}
\label{sec2}
In this section, we describe the general framework for our study of 
decoherence. We consider one- and two-qubit systems, coupled to an environment
which is modelled using random matrix theory. The one-qubit case is introduced 
in \sref{G1}, the two-qubit case in \sref{sec:twoqubitmodel}.
The quantities studied in this paper are defined in \sref{sec:themeas}
and the tools used to solve the problem are introduced in \sref{sec:generalprogram}.

\subsection{The model for one qubit}\label{G1} 

We describe decoherence by considering explicitly the additional degrees of
freedom (henceforth called ``environment'') which are interacting with the 
qubit. We follow the unitary time evolution of a pure, initially separable, 
state in the product space $\mcH= \mcH_1 \otimes \mcH_\e$, where $\mcH_1$ (of
dimension two) and $\mcH_\e$ (of dimension $N_\e$) denote the 
Hilbert spaces of the qubit and the environment, respectively. As time evolves, 
the qubit and the environment become more and more entangled, which means that 
after tracing out the environmental degrees of freedom, the state of the qubit 
becomes more and more mixed. For such a setup, the Hamiltonian is of the 
following form
\begin{equation}
  \label{eq:hamiltonianqubit}
H_\lambda= H_1\otimes \openone_\e + \openone_1 \otimes H_\e 
   + \lambda V_{1,\e} \equiv H_1 + H_\e + \lambda V_{1,\e} \; .
\end{equation}
Here, $H_1$  represents the Hamiltonian acting on the qubit, $H_\e$ the 
Hamiltonian of the environment, and $V_{1,\e}$ the coupling between the qubit 
and the environment. The real parameter $\lambda$ controls the strength of the 
coupling. For notational ease, and if there is no danger of confusion, we omit 
extensions with the identity, as it is done in \eref{eq:hamiltonianqubit}. 

We describe both the coupling and the dynamics in the environment within random
matrix theory. To this end, $H_\e$ and $V_{\e,1}$ are chosen from either the
GUE or the GOE, depending on whether we wish to describe a TRI 
breaking or TRI conserving situation. The Hamiltonian $H_1$ implies
another free parameter of the model, namely the level splitting $\Delta$ of the 
two level system representing the qubit.

We shall study the time evolution of an initially pure and separable state
\begin{equation}
  \label{eq:initialonequbit}
  |\psi(t=0)\> = |\psi_1\>\otimes |\psi_\e\> \; ,
\end{equation}
where $|\psi_1\> \in \mcH_1$ and $|\psi_\e\>\in \mcH_\e$. At any time $t$,
the state of the whole system is thus $|\psi(t)\>=\exp(-\mimath t H_\lambda)
|\psi(0)\>$, and the state of the single qubit is $\tr_\e |\psi(t)\>
\<\psi(t)|$.  While the state of the qubit $|\psi_1\>$ is yet another free
parameter in our model, we assume the state of the environment $|\psi_\e\>$ to
be random.  This means that the state is chosen from an ensemble which is
invariant under unitary transformations. In practice, this means that the
coefficients are chosen as complex random Gaussian variables, and subsequently
the state is normalized.

\subsection{The model for two qubits}\label{sec:twoqubitmodel}

For the two qubit case, we need to introduce the additional Hilbert spaces
$\mcH_2$ of dimension two and $\mcH_{\e'}$ of dimension $N_{\e'}$, which 
correspond to the second qubit and its environment respectively.
As in the one-qubit case, the dynamics in the space of the two qubits 
$\mcH_1\otimes\mcH_2$ (henceforth called ``central system'') is obtained by 
considering the unitary evolution in whole Hilbert space 
$\mcH_1\otimes\mcH_2\otimes\mcH_\e\otimes\mcH_{\e'}$ and subsequently 
tracing-out the environmental degrees of freedom (henceforth called
``environment''). As initial states, we use pure states in the whole Hilbert
space, which are separable with respect to any combination of 
$\mcH_1\otimes\mcH_2$, $\mcH_\e$, and $\mcH_{\e'}$:
\begin{equation}
  \label{eq:initialconditiongeneral}\fl\qquad
  |\psi(0)\>=|\psi_{12}\>|\psi_\e\>|\psi_{\e'}\>,\quad        |\psi_{12}\>\in
   \mcH_1 \otimes \mcH_2,\,|\psi_\e\>\in \mcH_\e,\,{\rm and}\,
  |\psi_{\e'}\>\in \mcH_{\e'}\; . 
\end{equation}
Typically, we assume that at $t=0$ the two qubits (and hence the state
$|\psi_{12}\>$) to be entangled. This is the main new ingredient, as compared 
to the one-qubit case. In this work we shall consider three different 
dynamical scenarios, all explicitly excluding any interaction between the 
two qubits:

\begin{figure}[thbp]
  \centering
  \includegraphics[width=\textwidth]{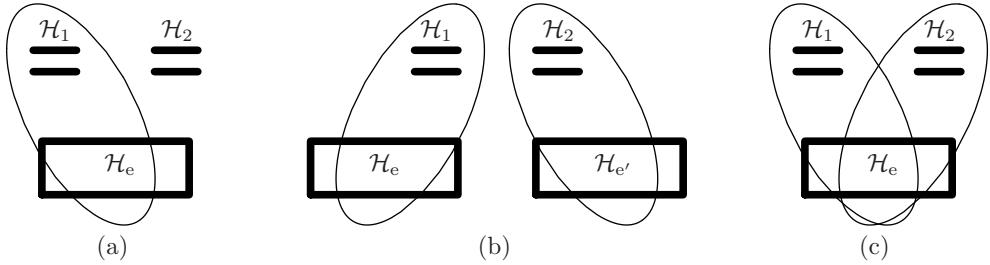} 
  \caption{Schematic representations  of the different dynamical 
    configurations studied in this article:  (a) the spectator Hamiltonian,
    (b) the separate, and (c) the joint environment Hamiltonian.}
  \label{fig:schemeconfig}
\end{figure}

\begin{itemize}
\item[(a)] {\em The spectator Hamiltonian:} 
In this case the two qubits evolve under separate Hamiltonians $H_1$ and $H_2$
acting on the Hilbert spaces $\mcH_1$ and $\mcH_2$, respectively.
In addition, we assume that only the first 
qubit is coupled to an environment, such that the total Hamiltonian reads
\begin{equation}\label{eq:spectatorHam}
H_\lambda= H_1 + H_2  + H_\e  + \lambda V_{1,\e} \; ,
\end{equation}
where $\lambda$, $H_1$, $H_\e$ and $V_{1,\e}$ are defined as in 
\eref{eq:hamiltonianqubit}.
This situation is shown schematically in \fref{fig:schemeconfig}(a). If we
choose an initial state where the two qubits are already entangled, this 
provides the {\it simplest} situation which allows to study entanglement decay. If
the two qubits are initially not entangled, the process reduces effectively
to the one-qubit case, described previously \sref{G1}. The special case
$H_1=H_2=0$ has been considered in Ref.~\cite{pinedaRMTshort}.

\item[(b)] {\em The separate environment Hamiltonian:} 
This case differs from the spectator model only inasmuch as the second qubit 
is also coupled to an environment. Both environments are assumed to be 
non-interacting. The total Hamiltonian then reads
\begin{equation}\label{eq:sepHam}
H_{\lambda_1,\lambda_2}= H_1+H_2+H_\e+H_{\e'}+\lambda_1\; V_{1,\e}
   + \lambda_2\; V_{2,\e'} \; ,
\end{equation}
where $V_{2,\e'}$ and $H_{\e'}$ describe the coupling to -- and the dynamics
in the additional environment. Both quantities are chosen independently from
the respective random matrix ensembles, in perfect analogy with $V_{1,\e}$ and
$H_\e$. The real parameters $\lambda_1$ and $\lambda_2$ fix the coupling
strengths to either environment. This model, see \fref{fig:schemeconfig}(b), 
may describe two qubits that are ready to perform a distant teleportation,
where each of them is interacting only with its immediate surroundings. It can 
also represent a pair of qubits that, although close together, interact with 
different and independent degrees of freedom.

\item[(c)] {\em The joint environment Hamiltonian:} 
The third case, shown in \fref{fig:schemeconfig}(c), describes a situation in
which both qubits are coupled to the same environment, even though the coupling
matrices are still independent. In that case, the total Hamiltonian reads
\begin{equation}\label{eq:joinedHam}
H_{\lambda_1,\lambda_2}= H_1+H_2+H_\e+\lambda_1\; V_{1,\e}
   + \lambda_2\; V_{2,\e} \; ,
\end{equation}
where $V_{2,\e}$ describes the coupling of the second qubit to the environment.
It is chosen independently from the same random matrix ensemble as $V_{1,\e}$.

\end{itemize}

\subsection{The measures}\label{sec:themeas}

As a measure of decoherence we use purity. This measures the degree of
mixedness of a density matrix $\rho$ representing a physical system. Purity is
defined as
\begin{equation}
  \label{eq:defpurity}
  P[\rho]= \tr \rho^2.
\end{equation}
It reaches a maximum value of one when the state is pure, \ie  when
$\rho=|\psi\>\<\psi|$ for some $|\psi\>$. Otherwise it is less than one and
reaches a minimum when the state is completely mixed. We use purity instead of
\eg the von Neumann entropy because it is simpler to handle from an algebraic 
point of view. Note that for one qubit both quantities are equivalent, and for 
two qubits they contain similar information~\cite{wei:022110}.

The other measure considered here is concurrence. It quantifies the degree of
entanglement of two qubits in a pure or mixed state. It is a measure used
extensively in the literature~\cite{andrereview} that is straightforward to 
calculate. It is closely related to the entanglement of formation, which 
measures the minimum number of Bell pairs needed to create an ensemble of pure states 
representing the state to be studied~\cite{firstconcurrence}.
Given a density matrix $\rho$ representing the state of two qubits,
concurrence is defined as
\begin{equation}
  \label{eq:concurrence}
  C(\rho)=\max \{0,\Lambda_1-\Lambda_2-\Lambda_3-\Lambda_4 \}
\end{equation}
where $\Lambda_i$ are the eigenvalues of the matrix $\sqrt{\rho (\sigma_y
  \otimes \sigma_y) \rho^* (\sigma_y \otimes \sigma_y)}$ in non-increasing
order. The superscript $^*$ denotes complex conjugation in the computational 
basis and $\sigma_y$ is a Pauli matrix~\cite{wootters}. The concurrence has a 
maximum value of one for Bell states, and a minimum value of zero for separable 
states. Furthermore, it is invariant under bilocal unitary operations.

Both, purity and concurrence, provide a measure of entanglement, but 
in the present context they quantify completely different aspects of the 
problem. Purity will be used to quantify the entanglement between the pair of 
qubits and the environment \ie decoherence. Concurrence will be used to measure 
the entanglement within the pair. Although they are not fully independent,  
information about one provides little knowledge about the other. \Eg a state 
with purity one can have full entanglement (a Bell pair) or null entanglement 
(a separable pure state).

\subsection{Echo dynamics and linear response theory}\label{sec:generalprogram}

We shall calculate the value of purity as a function of time analytically,
in a perturbative approximation. To perform this task it is useful to
consider each of the above Hamiltonians [\eref{eq:hamiltonianqubit},
\eref{eq:spectatorHam}, \eref{eq:sepHam}, and \eref{eq:joinedHam}],
as composed by an unperturbed part $H_0$ and a perturbation
$\lambda V$ with $\lambda$ a parameter.  The unperturbed part corresponds to
the operators that act on each individual subspace alone whereas the
perturbation corresponds to the coupling among the different subspaces; for
example in the one qubit case $H_0=H_\e+H_1$ and $V=V_{\e,1}$. For the case of
two qubits, $\lambda$ can be taken as $\max\{\lambda_1,\lambda_2\}$.
We shall use the tools developed in \ref{sec:aA} for a linear response
formalism in echo dynamics. In order to do so we must formally write the
problem in this language.

We write the Hamiltonian as
\begin{equation} \label{eq:hlambda}
  H_\lambda=H_0+\lambda V \; ,
\end{equation}
and introduce the evolution operator and the echo operator defined by
\begin{equation}  \label{eq:defevecho}
  U_\lambda(t)=e^{-\mimath H_\lambda t}\; , \quad
  M_\lambda(t)=U_0(t)^\dagger U_\lambda(t) \; ,
\end{equation}
respectively ($\hbar =1$).  For the calculation of purity at a given time $t$, we replace the
forward evolution operator $U_\lambda$ by the corresponding echo operator
$M_\lambda$. Even though the resulting states are different, \ie
$\rho(t)= \tr_{\e,\e'} U_\lambda(t) \rho U_\lambda^\dagger(t) \ne \rho^M(t)=
 \tr_{\e,\e'} M_\lambda(t) \rho M_\lambda^\dagger(t)$, they are still related by the
local (in the two qubits and the environment) unitary transformation $U_0(t)$.
Since such transformations do not change the entanglement properties, it holds
that $P(t)= P[\rho(t)]= P[\rho^M(t)]$. This step is crucial, since
the echo operator admits a series expansion with much larger range of validity
(both, in time and perturbation strength). The numerical simulations are all
done with forward evolution alone as they require less computational effort.

The Born expansion of the echo operator up to second order reads
\begin{equation} \label{eq:bornexpansion}
  M_\lambda(t)= \openone -\mimath \lambda I(t) - \lambda^2 J(t)
     + \Or(\lambda^3) \; ,
\end{equation}
with
\begin{equation}
  I(t)= \int_0^t \rmd \tau \tilde V(\tau), \quad
  J(t)= \intoh \tilde V(\tau) \tilde V(\tau')
\end{equation}
and $\tilde{V}(t)=U_0(t)^\dagger V U_0(t)$ being the coupling in the
interaction picture. Using this expansion we calculate the purity of the
central system, averaged over the coupling and the Hamiltonian of the
environment.

\section{One qubit decoherence}\label{O} 

We first study the GUE case with and without an internal Hamiltonian governing
the qubit.  The next step is to work out the GOE case.  There we concentrate on
the case with no internal Hamiltonian governing in the qubit since we want to
keep the discussion as simple as possible to focus on the consequences of the
weaker invariance properties of the ensemble. 

Here, we study the evolution under the Hamiltonian
\begin{equation}\label{eq:otrodos}
  H_\lambda= H_1 + H_\e + \lambda\; V_{1,\e} = H_0 + \lambda\; V_{1,\e} \; ,
\end{equation}
where the initial state
\begin{equation}
  \varrho_0= |\psi_1\> \<\psi_1| \otimes |\psi_\e\> \<\psi_\e| \; ,
\end{equation}
is the product of a fixed pure state $|\psi_1\> \in \mcH_1$ and a random pure
state $|\psi_\e\> \in \mcH_\e$; see \eref{eq:initialonequbit}. At any 
later time, the state of the qubit and its purity are given by
\begin{equation}
\rho(t)= \tr_\e\; U_\lambda(t)\, \varrho_0\, U_\lambda(t)^\dagger\qquad
P(t)= \tr_1\, \rho(t)^2 \; .
\label{eq:onequbitrho}
\end{equation}
In~\ref{sec:aA}, we compute the average purity $\< P(t)\>$ as a function of
time in the linear response approximation \eref{eq:bornexpansion}, following the steps outlined in
\sref{sec:generalprogram}. The average is taken with respect to the 
coupling $V_{1,\e}$, the random initial state $|\psi_\e\>$ and the spectrum of
$H_\e$. In the limit of $N_\e\to\infty$, 
we obtain~[\eref{B:pulrdef}, \eref{B:pulrires}]
\begin{eqnarray}
\fl\quad \< P(t)\> = 1 -2\,\lambda^2
   \int_0^t\rmd\tau\int_0^t\rmd\tau'\; 
   {\rm Re}\, A_{\rm JI}(\tau,\tau') + \Or(\lambda^4) \; ,
\label{eq:purityoneq}\\
\fl\quad A_{\rm JI}(\tau,\tau')=
   [C_1(|\tau-\tau'|)-S_1(\tau-\tau')]\bar C(|\tau-\tau'|)
  +\chi_{\rm GOE} [1-S'_1(-\tau-\tau')] \; ,
\label{eq:resajiuno}
\end{eqnarray}
where 
$\chi_{\rm GOE} = 1$ for the TRI case, and $\chi_{\rm GOE} = 0$ for the non-TRI
case. The correlation functions $C_1(\tau), S_1(\tau), S'_1(\tau)$, and 
$\bar C(\tau)$ are defined in~\ref{aB}. $\bar C(\tau)$ deserves special 
attention, since all the dependence of the environment is via the function
\begin{equation}\label{eq:thecorrelation}
\frac{1}{N_\e}\left\< \left| 
   {\textstyle\sum_{j=1}^{N_\e}}\rme^{-\mimath E_j t} \right|^2\right\>
 = \bar C(t) = 1 + \delta(t/\tau_H) - b_2^{(\beta)}(t/\tau_H) \; ,
\end{equation}
where the $E_j$'s are the eigenenergies of $H_\e$ and $\tau_H$ is the 
corresponding Heisenberg time. The two-point form factor, 
$b_2^{(\beta)}(\tau)$, is known analytically for the GUE ($\beta=2$) and the
GOE ($\beta=1$), which are the two cases treated here~\cite{mehta}.

\subsection{The GUE case}
\label{sec:gueone}

We are now in the position to give an explicit formula for $\< P(t)\>$ in the 
GUE case. This formula will generally depend on some properties of the initial 
condition $|\psi_1\>$. We wish to write $|\psi_1\>$ in the most general way, 
but grouping together cases, equivalent due to the invariance properties of the 
problem. 

Recall that $H=H_\e+H_1+ \lambda V$ represents an ensemble of Hamiltonians in 
which $H_\e \in \rm{GUE}$  and $V\in \rm{GUE}$, whereas $H_1$ (together with 
the initial condition $|\psi_1\>$) remains fixed throughout the calculation. 
The operations under which the ensemble is invariant are local (with respect
to the partitioning of the Hilbert space into $\mcH_1$ and $\mcH_\e$),
unitary (due to the invariance properties of the GUE), and leave $H_1$ 
invariant. Hence the transformation matrices must be of the form
$U\otimes \exp(\mimath \alpha H_1)$ with $\alpha$ a real number and $U$ 
a unitary operator acting on $\mcH_\e$. 

This freedom allows to choose a convenient basis to solve the problem. On
one hand, it allows to write $H_0$ in diagonal form (as done in \ref{sec:aA}), 
and on the other hand, we can use it to represent the initial state of the qubit
in such a way that there is no phase shift between the two components of the 
qubit. This can be achieved by appropriately choosing $\alpha$ (see also the 
discussion in \sref{sec:spectator}). We thus write, without loosing 
generality
\begin{equation}\label{eq:initialOneGUE}
	|\psi_1\>=\cos\phi |0\>+ \sin\phi |1\> \; ,
\end{equation}
where $|0\>$ and $|1\>$ denote the eigenstates of $H_1$. Notice
that if $\phi=0$ and $\phi=\pi/2$, $|\psi_1$ is an eigenstate of $H_1$. 
Finally, we choose
the origin of the energy scale in such a way that the Hamiltonian of the 
qubit can be written as $H_1=(\Delta/2)|0\>\<0|-(\Delta/2)|1\>\<1|$. Hence,
$\phi$ is the only relevant parameter describing the initial state.

We obtain the average purity from the general expression in 
\eref{eq:purityoneq} and \eref{eq:resajiuno}. For a pure initial state
$\rho_1= |\psi_1\> \<\psi_1|$ the relevant correlation functions 
${\rm Re}\, C_1(\tau), S_1(\tau)$ and $\bar C(\tau)$ are given in 
\eref{aB:ReC1}, \eref{aB:S1pure}, and \eref{B:Cbardef}, respectively. Using the 
symmetry of the resulting integrand with respect to the exchange of $\tau$ and 
$\tau'$, we find
\begin{equation}\label{eq:genGUEone}
\fl\quad
\< P(t)\>= 1 -4\lambda^2\int_0^t\rmd\tau\int_0^\tau\rmd\tau'\; 
   \bar C(\tau')\;
   \big [\, 1- g_\phi\; (1- \cos\Delta\tau')\, \big ] +
   \Or(\lambda^4,N_{\rm e}^{-1})
\end{equation}
with
\begin{equation}\label{eq:defg}
g_\phi=\cos^4\phi+\sin^4\phi = \frac{3+\cos(4\phi)}{4}
\end{equation}
quantifying the ``distance'' between $\phi$ and the eigenbasis of $H_1$.
  
Let us consider following two limits for $H_1$. The ``degenerate limit'', 
where the level splitting $\Delta$ is much smaller than the mean level spacing
$d_\e= 2\pi/\tau_H$ of the environmental Hamiltonian, and the ``fast limit'', 
where the level splitting is much larger. In the latter case, the  internal 
evolution of the qubit is fast compared with the evolution in the environment.

The degenerate limit leads to the known formula~\cite{pinedaRMTshort}
\begin{equation}\label{eq:DegenerateOne}
  P_{\rm D}(t)=1-\lambda^2 f_{\tau_\H}(t),
\end{equation}
with
\begin{equation}\label{eq:deff}
f_{\tau_\H}(t)=
   \cases{2t\tau_\H + \frac{2t^3}{3\tau_\H} & if $0\leq t<\tau_\H$,\\
          2t^2+\frac{2\tau_\H^2}{3}& if $t\geq\tau_\H $}.
\end{equation}
The result does not depend on the initial state of the qubit.
Due to the degeneracy all states are eigenstates of $H_1$ and thus equivalent.
The leading term of the purity decay is linear before the Heisenberg time and
quadratic after the Heisenberg time. Similar features were already observed in 
fidelity decay and purity decay in other contexts \cite{reflosch}.

In the fast limit ($\Delta\gg d_\e$), purity is obtained from 
\eref{eq:genGUEone} by replacing $\cos\Delta\tau'$  with one when it is 
multiplied with the $\delta$ function [see \eref{eq:thecorrelation}], and with 
zero everywhere else. For finite $N_\e$ care must be taken, since we
are assuming Zeno time (which is given by the ``width of the 
$\delta$-function'') to be much smaller than all other time scales, such that
$\Delta \ll N_\e\, d_\e$. The resulting expression is
\begin{equation}\label{eq:FastOne}
  P_{\rm F}(t)=1-\lambda^2 [ (1-g_\phi)f_{\tau_\H}(t)+2 g_\phi t \tau_\H]\;.
\end{equation}
Typically (depending on the initial state), this formula again displays a
dominantly linear decay below the Heisenberg time, and a dominantly quadratic
decay above, similar to \eref{eq:DegenerateOne}.

\begin{figure}
\centering \includegraphics[width=.8\textwidth]{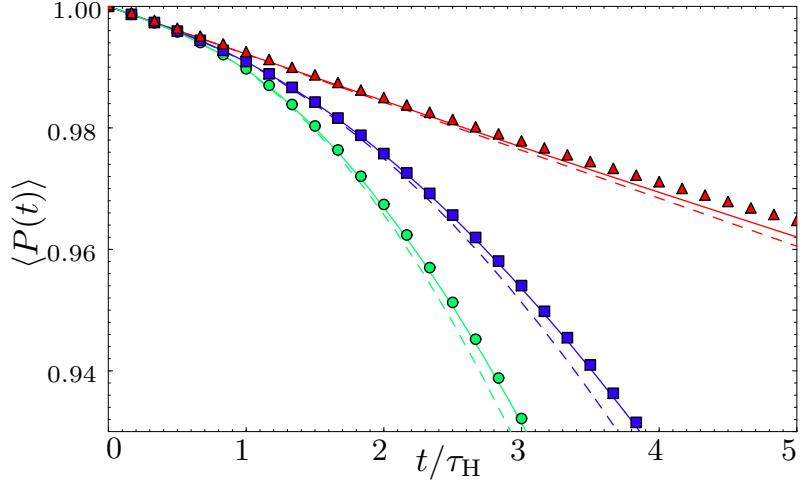}
\caption{
Numerical simulations for the average purity of one qubit using non-TRI
Hamiltonians, as a function of time in units
of the Heisenberg time $\tau_H$ of the environment. 
For the coupling strength 
$\lambda= 0.01$ and $N_\e=2048$, we show the dependence of $\< P(t)\>$ on $\Delta$ (the level 
splitting) and $\phi$ (characterizing the initial state) of the internal qubit:
$\Delta= 0, \phi= \pi/4$ (green circles), $\Delta= 8, \phi= \pi/4$ (blue 
squares), and $\Delta= 8, \phi= 0$ (red triangles). The corresponding linear
response results (dashed lines) and exponentiated linear response results
(solid lines) are based on \eref{eq:genGUEone} and \eref{eq:exponentiationone},
where $P_\infty$ is given in \ref{sec:exponentiation}. Note that the level 
splitting $\Delta$ tends to slow down decoherence.}
\label{fig:holeone}
\end{figure}

In \fref{fig:holeone} we compare numerical simulations of the average purity 
$\< P(t)\>$ (symbols) with the corresponding linear response result (dashed 
lines) based on \eref{eq:genGUEone}. The numerical results are obtained from
Monte Carlo simulations with 15 different Hamiltonians and 15 different initial 
conditions for each Hamiltonians. 
We wish to underline two aspects. First, the energy 
splitting in general leads to an attenuation of purity decay. Even though a 
strict inequality only holds for the limiting cases, 
$P_{\rm F}(t) > P_{\rm D}(t)$ (for $t\ne 0$), we may still say that 
increasing $\Delta$ tends to slow down purity decay. This result is in
agreement with earlier findings on  the stability of quantum 
dynamics~\cite{0305-4470-35-6-309}. Second, for the fast limit and an 
eigenstate of $H_1$ ($g_\phi=1$) we find linear decay even beyond the 
Heisenberg time. A similar behaviour has been obtained in~\cite{reflosch}, but
there an eigenstate of the whole Hamiltonian was required.

In \cite{pinedaRMTshort} it was shown that exponentiation of the linear 
response result leads to very good agreement beyond the validity of the 
original approximation. We use the formula \eref{eq:ELRextension}
\begin{equation}\label{eq:exponentiationone}
P_{\rm ELR}(t)=P_\infty+(1-P_\infty)
   \exp\left[- \frac{1-P_{\rm LR}(t)}{1-P_\infty}\right].
\end{equation}
where $P_{\rm LR}(t)$ is truncation to second order in $\lambda$ of the
expansion \eref{eq:genGUEone}, and $P_\infty=1/2$ the estimated
asymptotic value of purity for $t \to \infty$, see \ref{sec:exponentiation}. 
From \fref{fig:holeone} we see that the exponentiation indeed increases the 
accuracy of the bare linear response approximation.

\subsection{The GOE case}\label{sec:goeone}

\begin{figure}[htbp]
\centering
\includegraphics{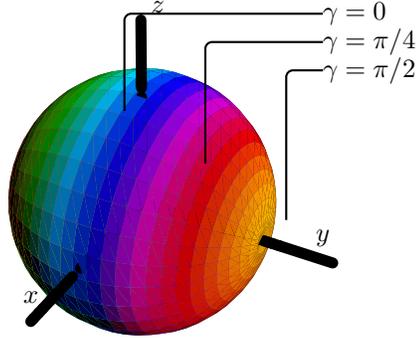} 
\caption{Any pure initial state of the qubit can be mapped onto the Bloch 
  sphere. Here, we show the angle $\gamma$ defined in \eref{eq:initialOneGOE}
  in colour code. Regions of a given colour represent subspaces which are
  invariant under the transformation $\exp(\mimath \alpha \sigma_y)$.}
\label{fig:blochgoe}
\end{figure}

We now drop $H_1$ leaving $H_0=H_\e$, resulting in 
\begin{equation}
	H_\lambda=H_\e+\lambda V
	\label{eq:noint}
\end{equation}
where $H_\e \in \rm{GOE}$ acts on $\mcH_\e$ and $V\in \rm{GOE}$ on
$\mcH_\e \otimes \mcH_1$. The resulting ensemble of Hamiltonians 
is invariant under local orthogonal transformations. In the environment, this 
allows again to diagonalize $H_\e$. In the qubit it allows rotations of the 
kind $\exp(\mimath \alpha \sigma_y)\in \mcO(2)$.  If such transformations are represented on the
Bloch sphere, they become rotations around the $y$ axis. Hence,
they can take any point on the Bloch sphere onto the $xy$-plane. Supposing this
point represents the initial state, it shows that we may assume $|\psi_1\>$ to
be of the form
\begin{equation}\label{eq:initialOneGOE}
	|\psi_1\>=\frac{|0\>+e^{\mimath \gamma}|1\>}{\sqrt 2}\; .
\end{equation}
In this expression, $\gamma\in [-\pi/2,\pi/2]$ denotes the angle of the point 
representing the initial state  with 
the $xz$-plane (see \fref{fig:blochgoe}).

In order to obtain the linear response expression for $\< P(t)\>$ we again
make use of \eref{eq:purityoneq} and \eref{eq:resajiuno}. However, apart from
the correlation functions used in the GUE case, we have now to consider in
addition $S'_1(\tau)$, as given in \eref{aB:S1ppure}. The special case
$H_1=0$ can simply be obtained by setting $\Delta =0$. This yields
\begin{equation} \label{goeone:aji}
        A_{\rm JI}(\tau,\tau')= \bar C(|\tau-\tau'|) + \sin^2\gamma \; .
\end{equation}
After evaluating the double integral in \eref{eq:purityoneq}, we obtain
\begin{equation} \label{eq:purityGOEsep}
 \< P(t)\>=1-\lambda^2 \left\{ t^2 \left[3-\cos(2\gamma)\right] +2 t \tau_H - 2 B_2^{(1)}(t)  \right\},
\end{equation}
where
\begin{equation} \label{eq:Btwogoe}
B_2^{(1)}(t)=  2\int_0^t \rmd \tau \int_0^\tau \rmd\tau'\; 
   b_2^{(1)}(\tau'/\tau_H) 
\end{equation}
is the double integral of the form factor. It can be computed analytically, but
the resulting expression is very involved~\cite{1367-2630-6-1-020}. For our
purpose it is sufficient to note that for 
$t \ll \tau_\H$,  $B_2^{(1)}(t) \propto t^3 $ (as in the GUE case), whereas
for $t \gg \tau_H$, $t- B_2^{(1)}(t)$ grows only logarithmically.  

\begin{figure}[htbp]
\centering
\includegraphics{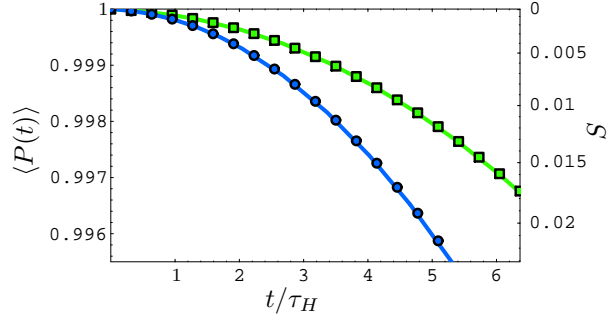}
\caption{The averages of the purity and the von Neumann entropy for initial
  states, ``located'' in different regions on the Bloch sphere. These regions
  are characterized by the angle $\gamma= 0$ (green squares) and 
  $\gamma= \pi/2$ (blue circles). The lines corresponding to the
  formula \eref{eq:purityGOEsep} with the appropriate value of $\gamma$. 
  Here, $N_\e=2048$ and $\lambda=10^{-3}$.}
\label{fig:decayGOE} 
\end{figure}

In \fref{fig:decayGOE} we show $\< P(t)\>$ for $\gamma= 0$ (green 
squares) and for $\gamma= \pi/2$ (blue circles).
In contrast to the GUE case (degenerate limit), the average purity depends on
the initial state via the angle $\gamma$. The fastest decay of purity is
observed for $\gamma= \pi/2$, where the image under the time reversal operation 
becomes orthogonal to the initial state. The slowest decay is observed for 
$\gamma= 0$, which characterizes states which remain unchanged under
the time reversal symmetry operation. 
These statements can be directly translated to the von Neumann entropy $S$.
For one qubit it has a one to one relation with purity
\begin{equation}\fl
        S(P)=h\left(\frac{1+\sqrt{2P-1}}{2}\right)+h\left(\frac{1-\sqrt{2P-1}}{2}\right),\,h(x)=-x\log_2x.
        \label{eq:entropur}
\end{equation}
Observe the entropy scale on~\fref{fig:decayGOE}.

Due to the dependence of \eref{eq:purityGOEsep} on $\gamma$, different
initial conditions in the Bloch sphere will yield different behaviours
of purity. However for a fixed value of $\gamma$ we numerically
check that there is self averaging (not shown), meaning that different members of the
ensemble behave the same as the average, in the large $N_\e$ limit.
As for one qubit the von Neumann entropy is a function of purity,
this behaviour will also be observed in this quantity. The consequences
of the weaker invariance properties of the GOE will be analysed in detail
in a more general framework in a latter paper.

\section{Two qubit decoherence}
\label{O2}

In this section, we address the question whether entanglement within a given
system affects its decoherence, if the system is coupled to an environment. We
discuss the three different situations described in \sref{sec:twoqubitmodel}.

\subsection{The spectator Hamiltonian \label{sec:spectator}}

The spectator Hamiltonian 
discussed in \sref{sec:twoqubitmodel}, reads
\begin{equation}\label{eq:effectivetwo}
H_\lambda= H_1+H_2+H_{\e}+\lambda V_{\e,1} \; .
\end{equation}
In order to study the decoherence of the initial state
\begin{equation}
\varrho_0= |\psi_{12}\>\, \<\psi_{12}| \otimes |\psi_{\rm e}\>\,\<\psi_{\rm e}| 
\; , 
\label{eq:initialcondtwoqu}
\end{equation}
we use the fact that the dynamics in $\mcH_2$ also
decouples from that in $\mcH_1$ and $\mcH_{\rm e}$, \ie that $H_2$ commutes
with all other terms in the Hamiltonian.
The quantum echo of  $\varrho_0$ after time $t$ is
\begin{equation}
\varrho^M(t)= \big (\, \openone_2\otimes M_\lambda(t)\, \big )\; \varrho_0\;  
   \big (\, \openone_2\otimes M^\dagger_\lambda(t)\, \big )\; .
	\label{eq:twoevolution}
\end{equation}
Since $\varrho^M(t)$ remains a pure state in 
$\mcH_1\otimes\mcH_2\otimes\mcH_{\rm e}$ for all times,
\begin{equation}
\fl\qquad P(t)= \tr \rho_{\rm c}(t)^2 = \tr \rho_\e(t)^2\qquad
\rho_{\rm c}(t)= \tr_\e \varrho^M(t)\qquad \rho_\e(t)= \tr_{\rm c} \varrho^M(t) \; .
	\label{eq:bothpuritiesequal}
\end{equation}
As the echo operator acts as the identity on the second qubit,
\begin{equation}
\fl\qquad
\rho_\e(t)= \tr_1  M_\lambda(t) ( \tr_2 \varrho_0 ) M^\dagger_\lambda(t) 
 = \tr_1  M_\lambda(t)\; \big (\, \rho_1\otimes|\psi_\e \>\<\psi_\e|\, \big )
   \; M^\dagger_\lambda(t) \; ,
\label{eq:ahora}
\end{equation}
where $\rho_1= \tr_2 |\psi_{12} \>\<\psi_{12}|$.
We may therefore compute the purity of the spectator model, without ever 
referring explicitly to the second qubit. Any dependence of the decay of the 
purity on the entanglement between the two qubits is encoded into the initial
density matrix $\rho_1$. This also implies that we can use the results obtained
in~\ref{sec:aA}, and hence \eref{eq:purityoneq} and \eref{eq:resajiuno} remain
valid. The only difference is that for the correlation functions $C_1(\tau)$, 
$S_1(\tau)$, and $S'_1(\tau)$, we now have to insert the respective expressions 
which apply for mixed initial states of the first qubit. These expressions are
given in~\ref{aB}.

\subsubsection{The GUE case:}\label{sec:guetwo}

We again wish to write the initial condition in its simplest form. We
must respect the structure of \eref{eq:spectatorHam}, but take advantage of all 
its invariance properties. Given a fixed $H_1$, the ensemble 
of Hamiltonians is invariant under local operations of the form
$U_{N_\e}\otimes\exp{\mimath \alpha H_1} \otimes U_2$ where
$U_{N_\e}\in\mcU(N_\e)$ is any unitary operator acting on the environment,
$\alpha$ a real number, and $U_2\in\mcU(2)$ is any unitary operator acting on 
the second qubit.

\begin{figure}[thbp]
\begin{center}
\includegraphics{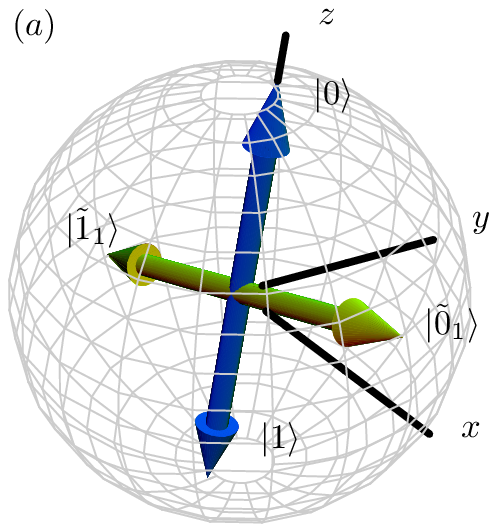}\hfill
   \includegraphics{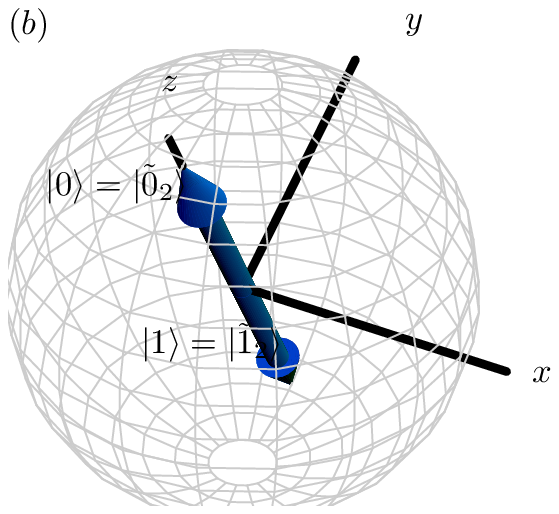}
\end{center} 
\caption{We see a plot to help visualize the way the initial condition
is parametrized. On the left, qubit one, has an internal Hamiltonian.
Its eigenvectors ($|0\>$ and $|1\>$) are represented in blue. The $z$
axis is {\textit chosen} parallel to the vector $|0\>$. The $x$ axis is chosen 
in such a way as to make both $|\tilde 0_1\>$ and $|\tilde 1_1\>$ have real
coefficients \ie such that the $xz$ plane contains $|\tilde 0_1\>$ and
$|\tilde 1_1\>$. On the right we represent the second qubit where we
have absolute freedom to choose the basis (even if an internal
Hamiltonian is present), and thus we choose it according to the natural
Schmidt decomposition.}
\label{fig:initGUEdos}
\end{figure}

The freedom in the environment allows to write $H_\e$ in diagonal from, 
whereas the freedom within the qubits allows to choose a basis 
$\{|0\>,|1\>\}\otimes \{|0\>,|1\>\}$ in which the state can be written as 
\begin{equation} \label{eq:initialguewitness} 
 |\psi_{12}\>=\cos\theta(\cos\phi|0\>+\sin\phi|1\>)|0\>+\sin\theta(\sin\phi|0\>-\cos\phi|1\>)|1\>,
\end{equation}
and still, $H_1=\frac{\Delta}{2}|0\>\<0|-\frac{\Delta}{2}|1\>\<1|$ is diagonal.
To find this basis we start using the Schmidt decomposition to write
\begin{equation}\label{eq:schmidt}
	|\psi_{12}\>=\cos\theta|\tilde 0_1 \tilde 0_2\> +  \sin\theta|\tilde 1_1
		\tilde 1_2\>
\end{equation}		
with $\{ |\tilde 0_i \>, |\tilde 1_i \> \}$ being an orthonormal
basis of particle $i$. For the first qubit, we fix the $z$ axis of the Bloch
sphere (containing both $|0\>$ and $|1\>$) parallel to the eigenvectors of $H_1$,
and the $y$ axis perpendicular (in the Bloch sphere) to both the $z$ axis and
$|\tilde 0_1 \>$. The states contained in the $xz$ plane are then real
superpositions of $|0\>$ and $|1\>$, which implies that $|\tilde 0_1
\>=\cos\phi|0\>+\sin\phi|1\>$ and $|\tilde 1_1 \> = \sin\phi|0\>-\cos\phi|1\>$
for some $\phi$.  In the second qubit it is enough to set $|0\>=|\tilde 0_2 \>$
and $|1\>=|\tilde 1_2 \>$. This freedom is also related to the fact that purity
only depends on $\tr_2 |\psi_{12}\> \<\psi_{12}|$. A visualization of this
procedure is found in \fref{fig:initGUEdos}. The angle $\theta \in [0,\pi/4]$ 
measures the entanglement ($C(|\psi_{12}\>\<\psi_{12}|) = \sin2\theta$) whereas
the angle $\phi \in [0,\pi/2]$ is related to an initial magnetization.

\begin{figure}[thbp]
  \centering
  \includegraphics[width=\textwidth]{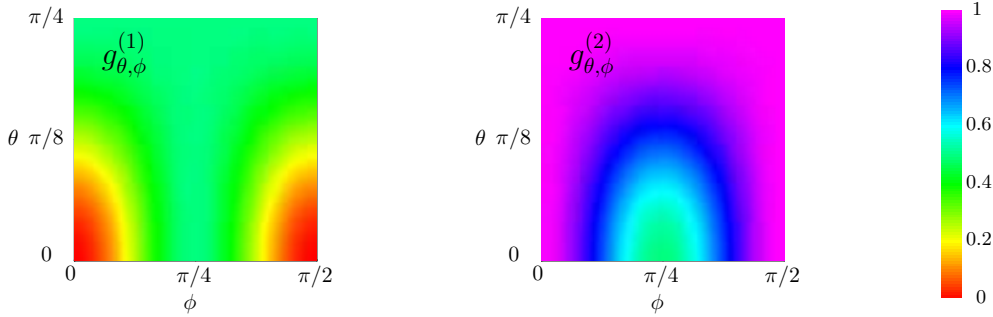}
  \caption{Visualization of the geometric factors
  $g_\theta$, $g_{\theta,\phi}^{(1)}$, and $g_{\theta,\phi}^{(2)}$ from left to
  right respectively. For $g_{\theta,\phi}^{(1)}$ we see that for pure
  eigenstates of $H_1$ its value is zero. This leads to a
  higher qualitative stability of this kind of states. }
  \label{fig:gsfig}
\end{figure}

The general solution for purity using this parametrization is
\begin{equation}\label{eq:generalGUEtwo}
\fl P(t)= 1 -4\lambda^2\int_0^t\rmd\tau\int_0^\tau\rmd\tau'\;
   \bar C(\tau')\; \big [\, g^{(1)}_{\theta,\phi} + 
   g^{(2)}_{\theta,\phi}\; \cos\Delta\tau'\, \big ] + 
   \Or(\lambda^4, N_{\rm e}^{-1})\; ,
\end{equation}
where the geometric factors $g^{(1)}_{\theta,\phi}\in [0,1/2]$, and 
$g^{(2)}_{\theta,\phi} \in [1/2,1]$  are expressed as
\begin{eqnarray}
g^{(1)}_{\theta,\phi}&= g_\theta (1-g_\phi) + g_\phi (1-g_\theta)\\
g^{(2)}_{\theta,\phi}&= 2(1-g_\theta) - g_\phi (1-2g_\theta),
\end{eqnarray}
in terms of the functions $g_\phi$ and $g_\theta$, defined in \eref{eq:defg}. Both geometric
factors are shown in \fref{fig:gsfig}. \Eref{eq:generalGUEtwo} is obtained
from \eref{eq:purityoneq} and \eref{eq:resajiuno} by insertion of the
\eref{aB:ReC1} and \eref{aB:S1mixed} for ${\rm Re}\, C_1(\tau)$ and 
$S_1(\tau)$, respectively.

We consider again two limits for $\Delta$. In the degenerate limit
($\Delta\ll 1/\tau_\H$) purity decay is given by
\begin{equation}\label{eq:spectatorDegenerate}
  P_{\rm D}(t)=1-\lambda^2(2-g_\theta)f_{\tau_\H}(t),
\end{equation}
where $f_{\tau_\H}(t)$ is defined in \eref{eq:deff}.
The result is independent of $\phi$ since a degenerate Hamiltonian
is, in this context, equivalent to no Hamiltonian at all.
The $\theta$-dependence in this formula shows that an entangled qubit pair is
more susceptible to decoherence than a separable one.

In the fast limit ($\Delta\gg 1/\tau_\H$) we get
\begin{equation}\label{eq:spectatorFast}
P_{\rm F}(t)=1-\lambda^2\left[g^{(1)}_{\theta,\phi}f_{\tau_\H}(t)
   + 2\tau_\H g^{(2)}_{\theta,\phi} t \right]\; .
\end{equation}
For initial states chosen as eigenstates of $H_1$ we find linear decay of 
purity both below and above Heisenberg time. In order for $\rho_1$ to be an 
eigenstate of $H_1$ it must, first of all, be a pure state (in $\mcH_1$). 
Therefore this behaviour can only occur if $\theta=0$ or $\theta= \pi/2$.
Apart from that particular case, we observe in both limits, the fast as well 
as the degenerate limit, the characteristic linear/quadratic behaviour 
before/after the Heisenberg time similar to the one qubit case.

\begin{figure}
\centering \includegraphics[width=.8\textwidth]{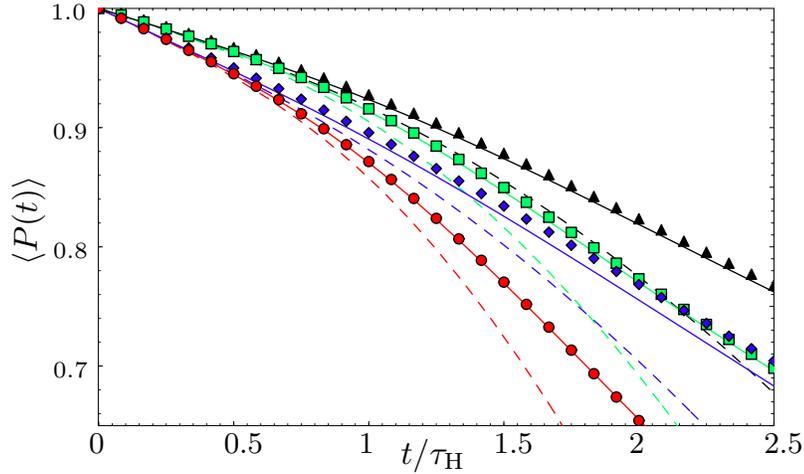}
\caption{
Numerical simulations for the average purity as a function of time in units of
the Heisenberg time of the environment (spectator configuration, GUE case).
For the coupling strength $\lambda= 0.03$ we show the dependence of 
$\< P(t)\>$ on the level splitting $\Delta$ in $H_1$ and on the initial degree 
of entanglement between the two qubits (in all cases $\phi= \pi/4$):
$\theta= 0$ (separable states), $\Delta=8$ (black triangles); $\theta= \pi/4$ 
(Bell states), $\Delta=8$ (blue rhombus); $\theta= 0$, $\Delta= 0$ (green 
squares); $\theta= \pi/4$, $\Delta= 0$  (red circles). The corresponding linear
response results (dashed lines and exponentiated linear response results (solid
lines) are based on \eref{eq:generalGUEtwo} and \eref{eq:exponentiationone}.
They are plotted with the same colour, as the respective numerical data.
In all cases $N_\e=1024$.}
\label{fig:holetwo}
\end{figure}

In \fref{fig:holetwo} we show numerical simulations for $\< P(t)\>$.
We average over 30 different Hamiltonians each 
probed with 45 different initial conditions. We contrast Bell states 
($\phi=\pi/4$, $\theta=\pi/4$) with separable states ($\phi=\pi/4$, $\theta= 0$), 
and also systems with a large level splitting ($\Delta= 8$) in the first qubit 
with systems having a degenerate Hamiltonian ($\Delta= 0$). The results presented
in this figure show that entanglement generally enhances decoherence. This can
be anticipated from \fref{fig:gsfig}, since for fixed $\phi$, increasing the 
value of $\theta$ (and hence entanglement) increases both 
$g^{(1)}_{\theta,\phi}$ and $g^{(2)}_{\theta,\phi}$. At the same time we find
again that increasing $\Delta$ tends to reduce the rate of decoherence, while
a strict inequality only holds among the two limiting cases (just as in the 
one qubit case). From
$g^{(2)}_{\theta,\phi}=2-g^{(1)}_{\theta,\phi}-g_\theta$, it follows that 
$(P_{\rm F}-P_{\rm D})/\lambda^2=g^{(2)}_{\theta,\phi} [f_{\tau_\H}(t)
  -2t\tau_\H] \ge 0$. Therefore, for fixed initial conditions and $t$ greater
than 0, $P_{\rm F}(t) > P_{\rm D}(t)$. This is the second aspect
illustrated in \fref{fig:holetwo}.

In order to extend the formulae to longer times/smaller purities we 
exponentiate them using the results of \ref{sec:exponentiation}. The numerical
simulations (see \fref{fig:holetwo}) agree very well with that heuristic 
exponentiated linear response formula. In one case (blue rhombus) where the 
agreement is not so perfect, we found that it is the inaccurate estimate
$P_\infty= 1/4$ of the asymptotic value of purity, which leads to the 
deviations.

\subsubsection{The GOE case:}\label{sec:goetwo}

Let us consider the GOE average of both  $H_\e$ and $V_{1,\e}$.  We have
to take into consideration the weaker symmetry properties of the GOE ensemble as
compared with the GUE.  When averaging $H_\e$ and $V_{1,\e}$ over the GOE, we
are again confronted with the fact that the invariance group is considerably
smaller than in the GUE case. In this context the initial entanglement between
the two qubits has a crucial importance since it ``transports'' the invariance
properties from the spectator to the coupled qubit.  

For the sake of simplicity, we focus on the degenerate limit setting $H_1=0$.
Note that on the basis of the results in~\ref{sec:aA}, the general case can be
treated similarly and the corresponding result will be presented at the end of 
this subsection.

We first specify the operations under which the spectator Hamiltonian 
\eref{eq:effectivetwo}, considered as a random matrix ensemble, is invariant. 
As both, the internal Hamiltonian of the environment and the coupling, are 
selected from the GOE the invariance operations form the group
\begin{equation} \label{eq:invariancegoe}
	\mcO (N_\e) \times \mcO(2) \times \mcU(2),
\end{equation}
and have the structure 
$O_{N_\e} \otimes \exp (\mimath \alpha \sigma_y) \otimes U_2$,
with $O_{N_\e}$ being an orthogonal matrix (acting in $\mcH_\e$), $\alpha$
a real number, and $U_2$ a unitary operator acting on the spectator qubit.

The direct product structure of the invariance group oblige us to respect the
identity of each particle, but allows to analyse each qubit separately. For
instance, if we would replace the random coupling matrix $V_{1,\e}$ with one, 
which involves both qubits, the invariance group would be 
$ \mcO (N_\e) \times \mcO(4)$. As a consequence, purity decay would become
independent of the entanglement within the qubit pair: 
for any entangled state one can find a orthogonal matrix which
maps the state onto a separable one.

We write the initial condition $|\psi_{12}\>$ as in
\eref{eq:schmidt}. For the coupled particle follows the same analysis made in
\sref{sec:goeone}.  We can thus write 
$|\tilde{0}_1\>=2^{-1/2}[|0\>+ \exp(\mimath \gamma)|1\>]$ and, in order to 
respect orthogonality,
$|\tilde{1}_1\>=2^{-1/2}\exp(\mimath \zeta)[|0\>- \exp(\mimath \gamma)|1\>]$.
For the second qubit we have the same complete freedom as in \eref{sec:guetwo}. 
We thus select $|\tilde{0}_2\>=|0\>$ and 
$|\tilde{1}_2\>=\exp(-\mimath \zeta)|1\>$ to erase the relative phase in
the first qubit and finally write the initial state as
\begin{equation}
	|\psi_{12}\>= \frac{\cos \theta (|0\>+ e^{\mimath \gamma} |1\>) |0\>
		 +\sin \theta  (|0\>- e^{\mimath \gamma} |1\>) |1\>}{\sqrt{2}}.
	\label{eq:psi12goe}
\end{equation}
The average purity is still given by the double integral expression in 
\eref{eq:purityoneq}. However, in the present case the mixed initial state
$\rho_1= {\rm tr}_2\, |\psi_{12}\> \<\psi_{12}|$ must be used. For $\Delta=0$, 
the resulting integrand reads
\begin{equation}
	A_{\rm JI}(\tau,\tau')= (2-g_\theta) \bar C(|\tau-\tau'|) 
	 + 1-g_\theta+ (2g_\theta -1)  \sin^2 \gamma ,
\end{equation}
where $C_1(\tau), S_1(\tau)$, and $S'_1(\tau)$ are given in
\eref{aB:ReC1}, \eref{aB:S1mixed}, and \eref{aB:S1pmixed}, respectively.
Evaluating the double integral, we obtain
\begin{eqnarray}
\label{eq:purityGOEone}\fl
\< P(t)\>=1-\lambda^2 \left\{ t^2 [4 - 2 \cos^2(2\theta)\cos^2 \gamma]
      +(4-2 g_\theta) \left[ t \tau_H - B_2^{(1)}(t) \right] 
   \right\} \; ,
\end{eqnarray}
where $B_2^{(1)}(t)$ is given in \eref{eq:Btwogoe}.
As in the GUE case, this result depends on the entanglement of the initial 
state, and, as in the one-qubit GOE case, it also depends on $\gamma$.  
Again it turns out that Bell states are more susceptible to decoherence than
separable states. Note however, that purity as a function of 
$\theta$ is not monotonous. Hence, a finite increase of entanglement does not
guarantee that the purity decreases everywhere in time. For separable states, 
$g_\theta= 1$, we retrieve formula \eref{eq:purityGOEsep}. However, for 
completely entangled states, $\theta= \pi/4$, and the dependence on $\gamma$ 
is lost. This is understood from a physical point of view, noticing
that any bilocal unital channel (in particular unitary operations) on a Bell
state can be reduced to a single local unitary operation acting on a single
qubit \ie for any local unitary operation $U_{1,2}$, there exists $U_1'$ such
that
\begin{equation}
  \label{eq:bilocaltolocal}
  U_1 \otimes U_2 |{\rm Bell}\>= U_1' \otimes \openone |{\rm Bell}\>
\end{equation}
($|\rm{Bell}\>$ is any 2 qubit pure state with $C=1$, e.g. $|00\>+|11\>$). 
We can then say that the invariance properties in the second qubit are
inherited from the first qubit via entanglement.

Let us obtain the standard deviation for the different possible initial conditions in
the qubits. We want to analyse the situation separately for a fixed value of
concurrence. Then, as the invariant measure of the ensemble of initial
conditions, and fixing the amount of entanglement, we shall use the tensor
product of the invariant measures in each of the qubits. Since there is no
dependence of \eref{eq:purityGOEone} on the second qubit, the appropriate invariant measure is
trivially inherited from the invariant measure for a single qubit. 
The resulting value for the standard deviation is
\begin{equation}
	\sigma_P=\frac{4 \lambda^2 t^2 \cos^2(2\theta)}{3 \sqrt{5}}.
	\label{eq:sigmados}
\end{equation}

Based on~\ref{sec:aA} and~\ref{aB} we can also obtain the average purity for
$\Delta\ne 0$. The parametrization of the initial states is more complicated
since two preferred directions arise, one from the eigenvectors of the internal
Hamiltonian and the other from the invariance group.  The result can be
expressed in the form given in \eref{eq:purityoneq}, with
\begin{eqnarray}
\fl\qquad {\rm Re}\, A_{\rm JI}(\tau,\tau')= 
   \bar C(|\tau-\tau'|)\; \big [\, g^{(1)}_{\theta,\phi} + 
   g^{(2)}_{\theta,\phi}\; \cos\Delta(\tau-\tau')\, \big ] \nonumber\\
\fl\qquad\qquad + g_{\theta,\phi}^{(1)} - (1-g_{\theta,\phi}^{(2)})\; 
   \cos[\Delta (\tau+\tau') -2\eta]  +
   \Or(\lambda^4, N_{\rm e}^{-1})\; . 
\end{eqnarray}
The angle $\eta$ is related to a phase shift between the components of any of
the eigenvectors of the initial density matrix $\rho_1$.

\subsection{The separate environment Hamiltonian}

We proceed to study purity decay with other configurations of the environment.
Consider the separate environment configuration, pictured in
\fref{fig:schemeconfig}(b).  The corresponding uncoupled Hamiltonian is
\begin{equation}\label{eq:sehamone}
	H_0 = H_1+H_2+H_{\e}+H_{\e'}
\end{equation}
and the coupling is 
\begin{equation}\label{eq:sehamdos}
	\lambda V=\lambda_1 V_{\e,1}+\lambda_2 V_{\e',2}. 
\end{equation}
From now on we assume that the internal Hamiltonians of the environment and the
couplings are chosen from the GUE. Generalization for the GOE can be obtained
along the same lines using the corresponding results of \sref{sec:goetwo}.
The initial condition has a separable structure with respect to both 
environments, see \eref{eq:initialconditiongeneral}. Next we calculate the
coupling in the interaction picture. It separates into two parts acting on
different subspaces
$\lambda \tilde V=\lambda_1 \tilde V^{(1)}+ \lambda_2 \tilde V^{(2)}$, where
\begin{equation}\label{eq:twopersep}
\fl\qquad
\tilde V^{(1)}=\rme^{\mimath (H_1 +H_{\e})} V_{\e,1} 
   \rme^{-\mimath (H_1 +H_{\e})},\quad
\tilde V^{(2)}=\rme^{\mimath (H_2 +H_{\e'})} V_{\e',2} 
   \rme^{-\mimath (H_2 +H_{\e'})}. 
\end{equation}
Notice that $\tilde V^{(1)}$ ($\tilde V^{(2)}$) does not depend on $H_{\e'}$ 
($H_{\e}$). Since $V^{(1)}$ and $V^{(2)}$ are uncorrelated quadratic averages 
separate as
\begin{equation}\label{eq:serpperint}
\lambda^2 \<\tilde V_{ij} \tilde V_{kl}\>  
   =\lambda_1^2 \<\tilde V^{(1)}_{ij} \tilde V^{(1)}_{kl}\>  
   +\lambda_2^2 \<\tilde V^{(2)}_{ij} \tilde V^{(2)}_{kl}\>.
\end{equation}
This leads to a natural separation of each of the contributions to
purity
\begin{equation}
1- \< P(t)\> = 1- P_{\rm spec}^{(1)}(t) + 1 - P_{\rm spec}^{(2)}(t) \; ,
\end{equation}
where $P_{\rm spec}^{(i)}(t)$ denotes the average purity with particle
$i$ being a spectator, as given in \sref{sec:spectator}. In this way, the problem
reduces to that of the spectator model. The respective expressions
in \sref{sec:spectator} may be used. For instance, if we assume broken 
TRI, we obtain from \eref{eq:generalGUEtwo}
\begin{equation} \label{eq:sepenvfull}
\fl
\< P(t)\>=1-4 \sum_{i=1}^{2} \lambda_i^2\intoh \left[g^{(1)}_{\theta,\phi_i}
  +g^{(2)}_{\theta,\phi_i}\cos\Delta_i\tau'\right] \bar C_i(\tau')
  + \Or(\lambda^4,N_\e^{-1}),
\end{equation}
where $\bar C_1$ and $\bar C_2$ are the correlation functions of the 
corresponding environments defined in exact correspondence with 
\eref{eq:thecorrelation}, for $H_\e$ and $H_{\e'}$ respectively. 
If in one or both of the qubits, the level splitting in the internal 
Hamiltonians is very large/small compared to the Heisenberg time in the
corresponding environment (denoted by $\tau_\e$ and $\tau_{\e'}$ for $H_\e$ and 
$H_{\e'}$, respectively) the degenerate and/or fast approximations may be used. 
As an example, if $\Delta_1 \ll 1/\tau_\e$ and $\Delta_2 \ll 1/\tau_{\e'}$ we 
find
\begin{equation}\label{eq:DeglimitSep}
P_{\rm D}(t)=1-(2-g_\theta)(\lambda_1^2 f_{\tau_\e}(t)+\lambda_2^2
  f_{\tau_{\e'}}(t)),
\end{equation}
whereas if $\Delta_1 \gg 1/\tau_\e$ and $\Delta_2 \gg 1/\tau_{\e'}$
\begin{equation}
  \label{eq:FastlimitSep} \fl\qquad
  P_{\rm F}(t)=1-
        \lambda_1^2\left[g^{(1)}_{\theta,\phi_1}f_{\tau_\e}(t)+2\tau_\e g^{(2)}_{\theta,\phi_1} t \right]
	- \lambda_2^2\left[g^{(1)}_{\theta,\phi_2}f_{\tau_{\e'}}(t)+2\tau_{\e'} g^{(2)}_{\theta,\phi_2} t \right].
\end{equation}
It is interesting to note that if we have two separate but equivalent 
environments
(\ie both Heisenberg times are equal), we get exactly the same result 
as for a single environment. 
Also notice that the
Hamiltonian of the entire system separates and thus the total
entanglement of the two subsystems ($\H_1 \otimes H_\e$
and $\H_2 \otimes H_{\e'}$) becomes time independent.

\subsection{The joint environment configuration}

The last configuration we shall consider is the one of
joint environment; see \fref{fig:schemeconfig}(c).
Its uncoupled Hamiltonian is
\begin{equation}\label{eq:johamu}
	H_0 = H_1+H_2+H_{\e}
\end{equation}
whereas the coupling is given by
\begin{equation}\label{eq:johamc}
	\lambda V=\lambda_1 V_{\e,1}+\lambda_2 V_{\e,2}. 
\end{equation}
Notice the similarity with \eref{eq:sehamone} and \eref{eq:sehamdos}. However, 
as discussed in the introduction, they represent very different physical 
situations. The coupling in the interaction picture can again be split 
$\lambda \tilde V=\lambda_1 \tilde V^{(1)}+ \lambda^{(2)} \tilde V_2$, where
\begin{equation}
\fl\qquad
\tilde V^{(1)}=\rme^{\mimath (H_1 +H_{\e})} V_{\e,1} \rme^{-\mimath (H_1 +H_{\e})},\quad
\tilde V^{(2)}=\rme^{\mimath (H_2 +H_{\e})} V_{\e,2} \rme^{-\mimath (H_2 +H_{\e})}. 
\end{equation}
Note the slight difference with \eref{eq:twopersep}. However still $V^{(1)}$ 
and $V^{(2)}$ are uncorrelated, enabling us to write again 
\eref{eq:serpperint}.

From  now  on,  the  calculation  is  formally the  same  as  in  the  separate
environment  case.  Hence  we  can  inherit  the  result  \eref{eq:sepenvfull}
directly, taking into  account that since they  come from the  same 
environmental Hamiltonian, the  two correlation functions are the same. In 
case any of the Hamiltonians fulfils the fast or degenerate limit conditions, 
the corresponding expressions to \eref{eq:DeglimitSep} and 
\eref{eq:FastlimitSep} can be written. As an example, if the first qubit  has 
no internal Hamiltonian, and the second  one has a big energy difference, the 
resulting expression for purity decay is
\begin{equation}\label{eq:joinedFast}
\fl\quad \< P(t)\> =1-\lambda_1^2 (2-g_\theta)f_{\tau_\H}(t)
  - \lambda_2^2\left[g^{(1)}_{\theta,\phi_2}f_{\tau_\H}(t)
  +2\tau_\H  g^{(2)}_{\theta,\phi_2} t \right]
  + \Or(\lambda^4,N_\e^{-1})\; ,
\end{equation}
where $\tau_\H$ is the Heisenberg time of the joint environment.  Monte Carlo
simulations  showing the  validity of  the result  were done  with satisfactory
results, comparable to those  obtained in \fref{fig:holetwo}. The parameter 
range checked was similar to that in the figure.

\section{A relation between concurrence and purity}
\label{R}

In the previous section we have found that purity decay is very sensitive to
the initial degree of entanglement between the two qubits. In other words, we 
related the behaviour of purity in time to the degree of entanglement (measured 
by concurrence) at $t=0$. In this section, we study the relation between purity 
and entanglement (again measured by concurrence) as both are evolving in time.
We focus on initial states which are Bell states such that purity and 
concurrence are both maximal and equal to one at $t=0$. Since concurrence is 
defined in terms of the eigenvalues of a hermitian 
$4\times4$-matrix, an analytical treatment even in linear response 
approximation is much more involved than in the case of purity. Instead, we 
work with a phenomenological relation between purity decay and concurrence 
discovered in \cite{pineda:012305} and further studied in 
\cite{pinedaRMTshort}. It proves to be valid in a wide parameter range, and it 
allows to obtain an analytic prediction for concurrence decay.

\begin{figure}[!ht]
\centering \includegraphics[height=4cm]{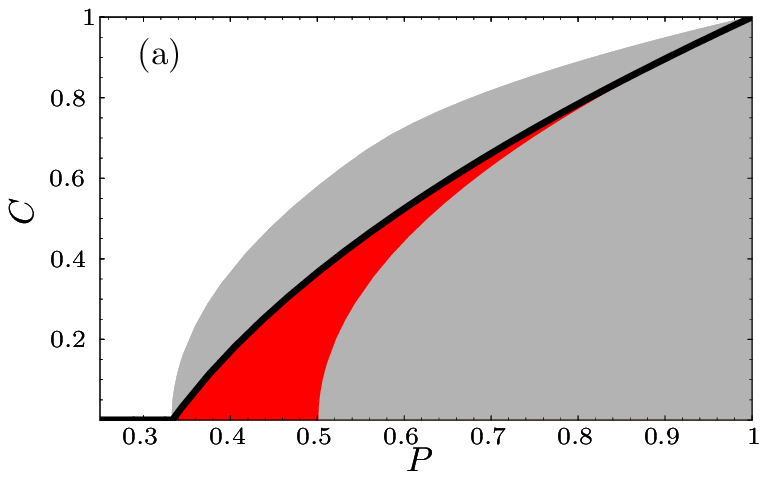}\hfill
\includegraphics[height=4cm]{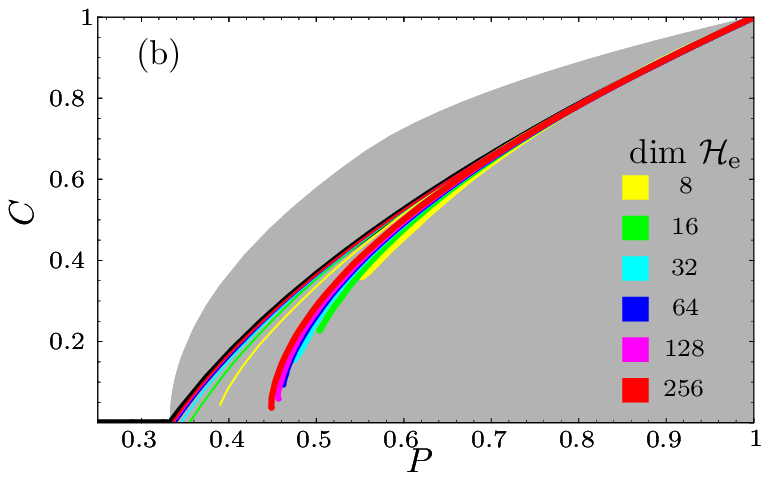}
\caption{
In (a) we show the area of concurrence-purity combinations which are allowed 
for physical states (grey area plus the set $\{(0,P),P\in[1/4,1/3]\}$). The
image of a Bell state under the set of bi-local unital operations defines the
red area. The Werner curve \eref{eq:goodexponwernerCtime}  is shown as a thick
black solid line. In (b) we show curves $(\<C(t)\>,\<P(t)\>)$ as obtained from
numerical simulations of the spectator Hamiltonian. GUE matrices 
are used during all numerical experiments in this section. We choose
$\Delta=1$ and vary $\dim(\mcH_\e)$ for two different coupling strengths
$\lambda=0.02$ (thin lines) and $\lambda=0.14$ (thick lines). The resulting
curves are plotted with different colours, according to $\dim(\mcH_\e)$ as
indicated in the figure legend.}
\label{fig:exampleacuuWerner_one}
\end{figure}

We study the relation between concurrence and purity on the $CP$-plane, where 
we plot concurrence against purity with time as a parameter. Since the initial 
state is a Bell state, the starting point is always at $(C,P)_{t=0}=(1,1)$. In 
this plane not all points are allowed for physical states; two constrains 
appear. The first one follows from the ranges of $C\in[0,1]$ and $P\in[1/4,1]$. 
The second restriction follows from the ``monogamy'' of entanglement: if a 
qubit is very entangled with another qubit, it cannot be very entangled with 
the rest of the universe (\ie the environment), implying some degree of purity. 
Conversely if the qubit is very entangled with the universe, it cannot be 
entangled with the other qubit. This statement can be made quantitative, which
leads to the concept of maximally entangled mixed states. They define the upper
bound of the set of admissible states (the corresponding area on the $CP$-plane
is shown in \fref{fig:exampleacuuWerner_one}(a) in grey). 
Another region of interest corresponds to those states, which form the image
of a Bell state under the set of bi-local unital operations 
\cite{ziman:052325} (red area in the same figure). Finally, we
have the Werner states 
$\rho_{\rm W}=\alpha \frac{\openone}{4}+ (1-\alpha)|\rm{Bell}\>\<\rm{Bell}|$, 
$0\le \alpha\le 1$, which define a smooth curve on the
$CP$-plane (black solid line).
The analytic form of this curve is \cite{pinedaRMTshort}
\begin{equation}\label{eq:goodexponwernerCtime}
C_{\rm W}(P)=\max\left\{0, \frac{\sqrt{12P-3}-1}{2}\right\},
\end{equation}
and will be referred to as the Werner curve. Note that states belonging to the 
Werner curve are not necessarily Werner states.

The spectator and separate environment configuration are strictly
bi-local, whereas the joint environment configuration becomes bi-local
in the large $N_\e$ limit. None of the three types of environmental couplings
are strictly unital. In this limit we find from our numerical simulations that all 
$CP$-curves fall into the region defined by the bi-local unital operations.

\begin{figure}[!ht]
\centering
\includegraphics[height=3.8cm]{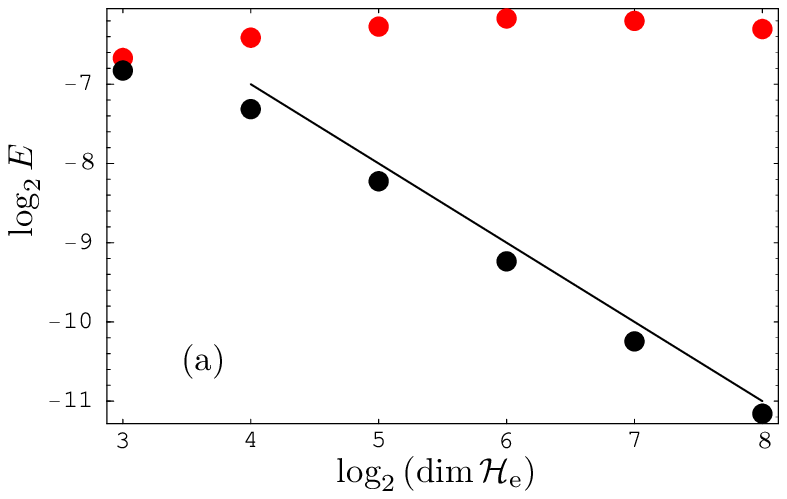}\hfill%
\includegraphics[height=3.8cm]{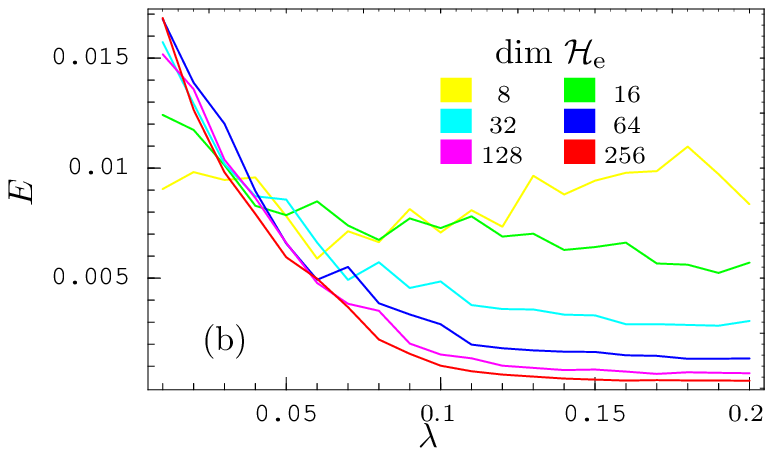}
\caption{
In (a) we show $E$ defined in \eref{eq:deferrorwerner} which measures the 
distance between the numerical $CP$-curves and the Werner curve as a function 
of $\dim N_\e$. The same data is used as in \fref{fig:exampleacuuWerner_one}.
The red dots refer to the case $\lambda=0.02$ where the error $E$ apparently
remains finite. The black dots refer to the case $\lambda=0.14$, where $E$
seems to tend to zero. The black solid line, which is proportional to 
$N_\e^{-1}$, is meant as a guide to the eye. In (b) we plot $E$ as a function of 
the coupling $\lambda$, for various values of $N_\e$. We may say that the line
$\lambda\approx 0.1$ separates two regimes. For $\lambda > 0.1$ we observe 
the convergence to the Werner curve, whereas for $\lambda < 0.1$ the limiting 
curve is a different one.}
\label{fig:transicion}
\end{figure}

For \fref{fig:exampleacuuWerner_one}(b), we perform numerical simulations of
the spectator Hamiltonian assuming broken time reversal symmetry (GUE case).
We compute the average purity $\< P(t)\>$ as well as the average concurrence
$\< C(t)\>$ as a function of time. This average is over 15 different 
Hamiltonians and 15 different initial states for each Hamiltonian. We fix the
level splitting in the coupled qubit to $\Delta=1$ and consider two different
values $\lambda=0.02$ and $0.14$ for the coupling to the environment.
\fref{fig:exampleacuuWerner_one}(b) shows the resulting $CP$-curves for 
different dimensions of $\mcH_\e$. 
Observe that for both values of $\lambda$, the curves converge to a 
certain limiting curve inside the red area (defined by the set of 
bi-local unital channels) as $\dim(\mcH_\e)$ tends to infinity. While for
$\lambda=0.02$, this curve is at a finite distance of the Werner curve, for
$\lambda=0.14$ it practically coincides with $C_{\rm W}(P)$.
To check this statement in more detail, consider the numerical $CP$-curve
$C_{\rm num}(P)$ obtained from our simulations and define its distance $E$ 
to the Werner curve as
\begin{equation}
  \label{eq:deferrorwerner}
  E=\int_{P_{\rm min}}^{1}\rmd P \left| C_{\rm num}(P)-C_{\rm W}(P)\right| \; .
\end{equation}
The behaviour of $E$ is shown in \fref{fig:transicion}(a). For $\lambda=0.14$
(black dots), the error goes to zero, which means that the corresponding 
$CP$-curves indeed converge to the Werner curve. In fact from a comparison with 
the black solid line we may conclude 
that the deviation $E$ is inversely proportional to the dimension of $\mcH_\e$.
By contrast, for $\lambda= 0.02$ (red dots), the error tends to an 
approximately constant value, in line with the assertion that the numerical 
results converge to a different curve.
In \fref{fig:transicion}(b) we plot the error $E$ as a function of $\lambda$,
for different dimensions of $\mcH_\e$ as indicated in the figure legend. The
results suggest that the convergence to the Werner curve occurs as long as 
$\lambda > \lambda_{\rm c}$ which is of the order of $0.1$ for $\Delta=1$. 
In fact for all $0.001 \le \Delta \le 100$ studied, we found a critical value 
$ \lambda_{\rm c}$ such that the $CP$-curves converge to the Werner curve as
long as $\lambda \gtrsim \lambda_{\rm c}$. For $\Delta=0$ all studied couplings 
led to convergence to the Werner curve.

In the presence of TRI
the $CP$-curves again converge to the Werner curve, for values of the coupling
greater than a critical value $\lambda_c$. As in the GUE case, this critical 
value depends on $\Delta$. However, in contrast to the GUE case, we find
that $\lambda_c$ remains finite for $\Delta=0$.
In the other configurations considered (the joint and the
separate environment), the behaviour is similar. In those cases 
it is the largest (of the two) coupling strengths which determines whether the 
$CP$-curves converge to the Werner curve, or not.

A partial explanation for this particular behaviour in the $CP$ plane is
provided in \cite{ziman:052325}. The authors find that for unital channels, the
possible points in the $CP$ plane that can be accessed are limited to a small
2-dimensional region, see \fref{fig:exampleacuuWerner_one}(a).
\begin{figure}
  \centering \includegraphics[width=.8\textwidth]{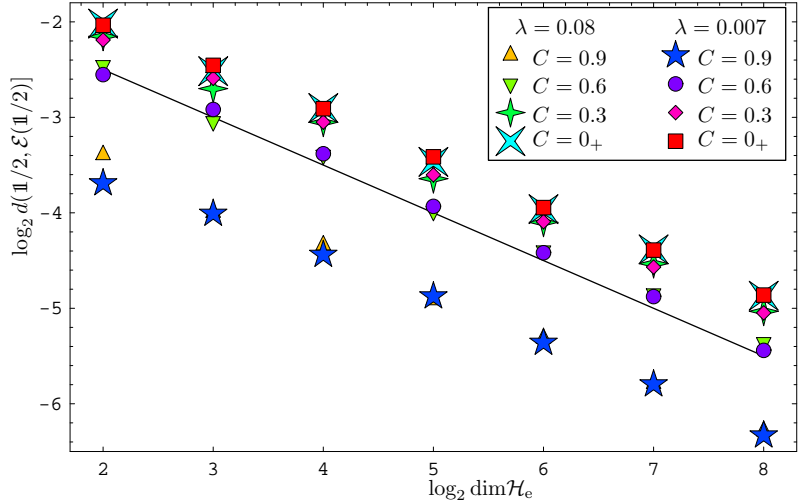}
  \caption{
  We evaluate the unitality condition for the spectator configuration (\ie
  on one of the qubits). We vary the size of the environment and test 
  different times, that would lead to different values of concurrence
  as shown in the inset, for the spectator case with the corresponding
  coupling. A line with slope $-1/2$ is also included for comparison.  
  }
  \label{fig:theplothatgorinwanted}
\end{figure}

We test the unitality of the channels induced by the non-unitary evolution
in the following way. Since we want to test $\mathcal{E}(\openone)=\openone$,
we prepare an initial pure condition that leads to a completely mixed state in
the qubit (let us focus on the spectator case), \ie,
\begin{equation}
\frac{|0 \psi_\e^{(0)} \> + |1 \psi_\e^{(1)} \>}{\sqrt{2}}
\end{equation}
with $\< \psi_\e^{(i)} | \psi_\e^{(j)} \>=\delta_{ij}$. We let the state
evolve with a particular member of the ensemble of Hamiltonians defined in
\eref{eq:hamiltonianqubit}. Afterwards we evaluate the Euclidean distance of
the resulting mixed state in the qubit from the fully mixed state, within the
Bloch sphere. The average distance is plotted as a function of the size of the
environment in \fref{fig:theplothatgorinwanted}.  We conclude that the
unitality condition is approached algebraically fast as the size of the
environment increases.  This by no means explains the generic accumulation to
the Werner curve, but certainly restricts the possibilities; furthermore, for
high purities/concurrences the area converges to the curve $C=P$, hence
providing a satisfactory explanation in this regime.

\begin{figure}[!ht]
\centering
\includegraphics[height=3.8cm]{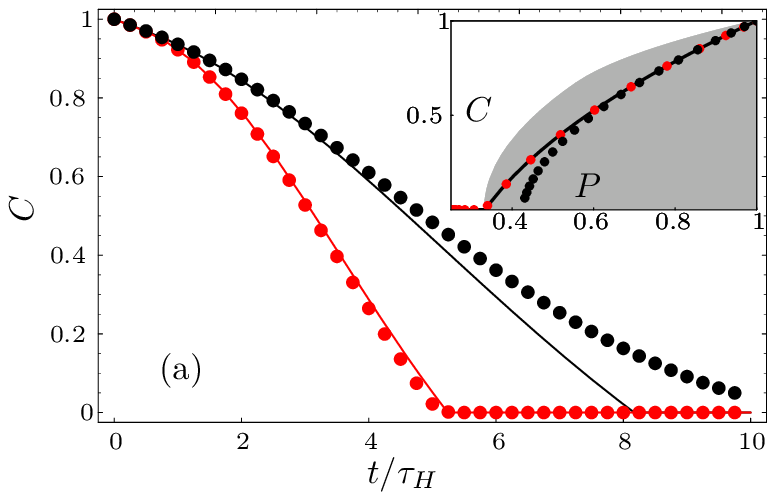}\hfill%
\includegraphics[height=3.8cm]{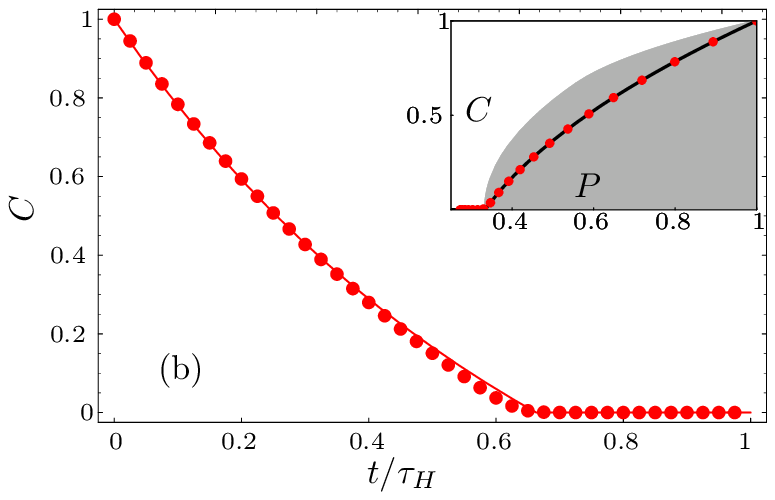}
\caption{
We show numerical simulations for the average concurrence as a function of
time in the joint environment configuration (GUE case). 
In (a) we consider small couplings $\lambda_1=\lambda_2=0.01$ which lead to the 
Gaussian regime for purity decay. The red symbols show the result without 
internal dynamics ($\Delta_1=\Delta_2=0$), whereas the black symbols are 
obtained for fast internal Rabi oscillations ($\Delta_1=\Delta_2=10$). The 
corresponding theoretical expectation based on \eref{eq:concurrenceELR} and 
\eref{eq:sepenvfull} is plotted as a red solid line (the case 
$\Delta_1=\Delta_2=0$), and a black solid line ($\Delta_1=\Delta_2=10$),
respectively. In (b) we consider stronger couplings, $\lambda_1=\lambda_2=0.1$,
such that purity decay becomes essentially exponential (Fermi golden rule
regime), while the level splitting have been set to $\Delta_1=\Delta_2=0.1$
(black symbols). The theoretical expectation (red solid line) is based on the
same expressions as in (a). The insets in (a) and (b) display the corresponding 
evolution in the $CP$-plane with the same symbols as used in the main graph.
In addition, the physically allowed region (grey area) and the Werner curve 
(black solid line) are shown. In all cases $N_\e=1024$}
\label{fig:CPdecay}
\end{figure}

Sufficiently close to $P=1$, the above arguments imply a one to one 
correspondence between purity and concurrence, which simply reads $C\approx P$.
This allows to write an approximate expression for the behaviour of 
concurrence as a function of time
\begin{equation}
  \label{eq:concurrenceLR}
  C_{\rm LR}(t)=P_{\rm LR}(t),
\end{equation}
using the appropriate linear response result for the purity decay. The 
corresponding expressions have been discussed in detail in the previous 
section. \eref{eq:concurrenceLR} has similar limits of validity as the linear
response result for the purity. Thus, in a sense, we may call it a linear
response expression for concurrence decay.

In those cases, where the coupling is beyond the critical coupling 
$\lambda_{\rm c}$ and where the exponentiated linear response expression
\eref{eq:exponentiationone} holds for the average purity, we can write down
a phenomenological formula for concurrence decay, which is valid over the 
whole range of the decay
\begin{equation}
  \label{eq:concurrenceELR}
  C_{\rm ELR}(t)=C_{\rm W}(P_{\rm ELR}(t)) \; .
\end{equation}
In \fref{fig:CPdecay} we show random matrix simulations for concurrence decay
in the joint environment configuration.
We consider small
couplings $\lambda_1=\lambda_2= 0.01$ which lead to the Gaussian regime for
purity decay, as well as strong couplings $\lambda_1=\lambda_2= 0.1$ which 
lead to the Fermi golden rule regime. We find good agreement with the 
prediction \eref{eq:concurrenceELR}, except for the Gaussian regime when we 
switch-on a fast internal dynamics in both qubits ($\Delta_1=\Delta_2= 10$).
These level splitting are so large, that the coupling strengths are much
smaller than $\lambda_{\rm c}$, which leads to the deviations observed. 

Note that, in the separate environment case, the entanglement of the two
subsystems defined on the spaces $\mcH_1\otimes\mcH_\e$ and 
$\mcH_2\otimes\mcH_{\e'}$ is constant in time. For the initial conditions we 
use, the entanglement stems entirely from the two qubits. The concurrence of
these two qubits decays because the constant entanglement spreads over all
constituents of the two large subspaces.

\section{Conclusions and outlook}\label{sec:conclusions}

We have derived general linear response formulae for the decay of purity 
for one and two qubits. The initial states are grouped in relevant families and
the dependence on these families is analysed in detail. In particular we see that 
time reversal invariance will generally increase the number of families 
we have to consider, though entanglement will undo some of this in the two-qubit case.
We do not discuss the well-known quadratic decay at very short (Zeno) times, though it 
obviously occurs in our model \cite{reflosch}. We find the linear decay
resulting from the usual master equation treatment at short times, but we see
the crossover to quadratic decay as we approach the Heisenberg time. This crossover is 
important if decoherence is dominated by a few degrees of freedom coupled to the
central system, leading to a finite Heisenberg time in the relevant environment.
We show that the quadratic term is absent if we choose an eigenstate 
of the intrinsic Hamiltonian of the central system. 

In general we can say that entanglement between the qubits accelerates decoherence
mainly by affecting the prefactors in the linear response expression. Internal 
dynamics introduces an additional term but also affects the prefactors.
It tends to stabilize the system though the interplay with entanglement does not make this 
statement strict for small changes.

The analysis of concurrence decay and its relation to purity decay largely
confirms previous findings \cite{pinedaRMTshort, pineda:012305} in this more
general setting. In particular the time-evolution follows the purity -
concurrence relation of Werner states.  We use this relation to
give an analytic, albeit heuristic, formula for concurrence decay.
Actually the relation shows small deviations for very weak couplings, and small
concurrences. We obtain some understanding of these findings by showing that
the process becomes unital in the limit of a large environment.

We thus have given a rather complete description of the time-evolution of
decoherence and entanglement of two non-interacting qubits. From a technical point of 
view we find a remarkable sequence of argumentation. 
In linear response approximation, all two-qubit configurations
can be reduced to the spectator case. This case is equivalent to
a single qubit, initially in a mixed state yet
not entangled with the environment. The evolution of the mixed central
system is obtained in the appendix in the linear response approximation. This
result in turn allows to derive the general formulae for all cases mentioned
above.
While this allowed a simple unified representation of the two-qubit problem in
the framework of an RMT model, actually one readily suspects, that this result is more general, 
and this is indeed the case as we shall see in a forthcoming paper \cite{GPS-letter}.
Another line, that could be easily followed, is to consider a random interaction between the two qubits and
a joint coupling to the environment. While such a study is easy and has some interest, the real problem 
to tackle is to introduce time dependent operations such as gates, 
while the decoherence modelled by
RMT acts. Another open problem that can be tackled in this framework is a situation of a 
near and far environment the first coupled weakly and the second extremely weakly.
The first should probably have a finite Heisenberg time on a time scale
relevant to the system, while the second more likely would have very dense 
spectrum and thus an infinite Heisenberg time. 
Finally, recall that we used formally echo dynamics for purely technical reasons.
Yet this could be used to great advantage if we simultaneously want to study
internal errors in the implementation of a quantum algorithm on a physical 
device as well as decoherence effects. 
As RMT is extremely successful in describing fidelity decay it 
seems to be the optimal approach. 

\ack
We thank Fran\c{c}ois Leyvraz, Walter Strunz, Peter Zoller, Toma\v z Prosen,
Andreas Buchleitner and Stefan Mossmann for many useful discussions.  We
acknowledge support from the grants UNAM-PAPIIT IN112507 and CONACyT 44020-F.
C.P. was supported by DGEP and VCCIC.

\appendix
\section[\\One-qubit purity decay for pure and mixed states]{One-qubit purity decay for pure and mixed states}
\label{sec:aA}

Here, we consider the one-qubit decoherence in a more general context.
One-qubit decoherence as described in \sref{O} deals with a separable 
initial state $\varrho_0= \rho_1 \otimes |\psi_{\rm e}\> \<\psi_{\rm e}|$, 
where $\rho_1= |\psi_1\> \<\psi_1|$ is a pure state of a single qubit coupled 
to an environment in the pure state $|\psi_{\rm e}\>$. We consider here
arbitrary, finite dimensions of $\mcH_1$ and $\mcH_{\rm e}$, and 
we allow $\rho_1$ to be mixed. The latter poses some problems on the physical
interpretation of purity as a measure for decoherence. However, these can be
avoided by taking the point of view of the environment, as it becomes entangled
with the central system. That means we consider the purity of the state of 
the environment after tracing out the central-system's degrees of freedom; 
see \eref{eq:ahora}. 
If we consider the entanglement with a spectator being responsible for the
mixedness of $\rho_1$, we can use the following results to describe purity decay in 
the spectator model.

In this sense, the result can be applied to the 
spectator model described in \sref{sec:spectator}.

We will derive explicit expressions for the average purity $\< P(t)\>$
as a function of time, where the average is only over the random coupling.
These expressions involve a few elemental spectral correlation functions, whose
properties will be discussed in \ref{aB}.
As detailed in \sref{sec:generalprogram}, we work with the echo operator
$M(t)$ in the linear response approximation \eref{eq:bornexpansion}. Note
that $I(t)$ is hermitian, while, due to time-order inversion, $J(t)$ is not.

\paragraph{The purity of the evolving state}
Let $\varrho_1$ and $\varrho_2$ be two density matrices, defined on the Hilbert 
space $\mcH'= \mcH_1\otimes\mcH_{\rm e}$. We define the 
purity form $p[\, \cdot\, ]$ as a function of pairs of such quantum states,
\begin{equation}\label{aA:defP}
  p[\varrho_1\otimes\varrho_2]= \tre[\tr_1(\varrho_1)\, \tr_1(\varrho_2)]\; .
\end{equation}
Since any linear operator acting on $\mcH' \otimes \mcH'$ can be expanded in
terms of separable operators of the form $\varrho_1\otimes\varrho_2$, 
linearity implicitly defines the purity form for arbitrary linear operators.
For arbitrary operators $A,B$ acting both on $\mcH'$, the purity form has the 
following property:
\begin{equation}\label{B:pfprop}
  p[A \otimes B] = p[A^\dagger \otimes B^\dagger]^* = p[B \otimes A] = p[B^\dagger \otimes A^\dagger]^* \; .
\end{equation}
The purity defined in \eref{eq:defpurity} can be expressed in terms of the purity form as
\begin{equation}
P(\tr_\e \varrho)= p[\varrho\otimes\varrho] \; .
\end{equation}
Note that on the RHS of this equation, we first take the trace over the 
central system, \ie we consider the purity of the state of the environment
after tracing out the central degrees of freedom. With this little twist we can
describe the one-qubit decoherence in \sref{O} and the spectator model in
\sref{sec:spectator} at the same time.

\paragraph{Purity decay in the linear response approximation}
To average purity, we use \eref{eq:bornexpansion} to compute
$\varrho^M(t)\otimes \varrho^M(t)$ in linear response approximation
\begin{equation}
\< P(t)\>= \<\, p[\varrho^M(t)\otimes\varrho^M(t)]\, \> 
         = p[\< \varrho^M(t)\otimes\varrho^M(t)\>] \; .
\end{equation}
Keeping terms only up to second order in $\lambda$:
\begin{eqnarray}
\fl \varrho^M(t)\otimes \varrho^M(t) = \varrho_0 \otimes \varrho_0 \nonumber\\
\fl  \qquad - \mimath\lambda \left[
   I\, \varrho_0 \otimes \varrho_0 - \varrho_0\, I^\dagger
   \otimes \varrho_0 + \varrho_0 \otimes I\, \varrho_0 - \varrho_0 \otimes
   \varrho_0\, I \right]^\dagger \\
\fl  \qquad - \lambda^2 \left[
   J\, \varrho_0 \otimes \varrho_0 + \varrho_0\, J^\dagger
   \otimes \varrho_0 + \varrho_0 \otimes J\, \varrho_0 + \varrho_0 \otimes
   \varrho_0\, J \right]^\dagger \nonumber\\
\fl \qquad + \lambda^2 \left[
   I\, \varrho_0\, I^\dagger \otimes \varrho_0 - I\, \varrho_0 \otimes I\,
   \varrho_0 + I\, \varrho_0 \otimes \varrho_0\, I^\dagger +
   \varrho_0\, I^\dagger \otimes I\, \varrho_0 - \varrho_0\, I^\dagger \otimes
   \varrho_0\, I^\dagger + \varrho_0 \otimes I\, \varrho_0 I^\dagger \right] 
   \; .
 \nonumber
\end{eqnarray}
In the next step, we perform the ensemble average over the coupling. As a
result, the terms linear in $\lambda$ vanish. For the remaining terms, we use
the properties in \eref{B:pfprop} and the fact that $I(t)$ is hermitian,
to obtain
\begin{eqnarray}\label{eq:pajai}
\fl\qquad \< P(t)\> = P(0) -\lambda^2\; (A_J - A_I) \qquad
P(0)= p[\varrho_0\otimes\varrho_0] \nonumber\\
\fl\qquad\qquad
   A_J = 4\, {\rm Re}\; p[\< J\> \, \varrho_0\otimes\varrho_0]\nonumber\\
\fl\qquad\qquad A_I = 2\big (\, p[\< I\, \varrho_0\, I\> \otimes \varrho_0]
   - {\rm Re}\, p[\< I\, \varrho_0 \otimes I\, \varrho_0\>]
   + p[\< I\, \varrho_0\otimes\varrho_0\, I\>] \, \big ) \; .
\end{eqnarray}
In order to pull out the same double integral of $A_J$ and $A_I$, we use the
time ordering symbol $\mathcal{T}$. It allows to write for the average purity
as a function of time
\begin{eqnarray}
\fl\qquad \< P(t)\>= P(0) -2\lambda^2\int_0^t\rmd\tau\int_0^t\rmd\tau'\; 
   {\rm Re}\, A_{\rm JI}\label{B:pulrdef}\\ 
\fl\qquad A_{\rm JI}= 
   p[\mathcal{T}\<\tilde V \tilde V'\> \varrho_0\otimes\varrho_0]
   - p[\<\tilde V' \varrho_0 \tilde V\> \otimes\varrho_0] 
   + p[\<\tilde V \varrho_0\otimes \tilde V' \varrho_0\>] 
\nonumber\\
\fl\qquad\qquad\qquad
   - p[\<\tilde V' \varrho_0\otimes \varrho_0 \tilde V\>] \; ,
\label{B:AJIdef}
\end{eqnarray}
where $\tilde V$ and $\tilde V'$ are short forms for the coupling matrices
$\tilde V(\tau)$ and $\tilde V(\tau')$, respectively. The arguments $\tau$
and $\tau'$ of the coupling matrices are interchanged in the second and fourth
term of $A_{\rm JI}$. This does not change the value of the integral of course,
but it facilitates to handle common terms in the following considerations.

\paragraph{The general result}
If the coupling matrix is taken either from the Gaussian
unitary (GUE) or orthogonal (GOE) ensemble, we find
\begin{eqnarray}
\fl\quad A_{\rm JI}= \big [\, C_1(|\tau-\tau'|) - S_1(\tau-\tau')\, \big ]\,
   \big [\, C_{\rm e}(|\tau-\tau'|) - S_{\rm e}(\tau-\tau')\, \big ] \nonumber\\
\fl\qquad\qquad + \chi_{\rm GOE}
   \big [\, 1 - S'_{\rm e}(-\tau-\tau') + S'_1(\tau+\tau')\;
   S'_{\rm e}(\tau+\tau') - S'_1(-\tau-\tau')\, \big ] \; .
\label{B:pulrres}
\end{eqnarray}
These expressions are derived in the following two subsections, for the GUE case 
in \ref{SU}, for the GOE case in \ref{sec:appGOE}. The correlation functions 
$C_x, S_x, C'_x$ and $S'_x$ with $x\in \{1,{\rm e}\}$ are defined and 
discussed in \ref{aB}. 

\paragraph{The limit $\dim(\mcH_{\rm e})\to\infty$:}
If the dimension of the Hilbert space of the environment goes
to infinity, the corresponding correlation functions simplify, as discussed 
in~\ref{aB}. We are then left with
\begin{equation}\label{B:pulrires} \fl\qquad
  A_{\rm JI}= \big [\, C_1(|\tau-\tau'|) - S_1(\tau-\tau')\, \big ]\;
     \bar C(|\tau-\tau'|) + \chi_{\rm GOE}\big [\, 1 - S'_1(-\tau-\tau')\, \big ] \; .
\end{equation}

\subsection{\label{SU}Appendix A.1 The GUE case}

The integrand $A_{\rm JI}$ in \eref{B:AJIdef} consists of four terms. 
These terms will be considered, one after the other. We will always first
average the argument of the purity form. Only then we perform the partial 
traces over subsystem $\mcH_1$, and therefore the final trace over the environment.
The averaging is only over the coupling, and it is done by applying two
simple rules, as described below.

For the coupling $V_{1,{\rm e}}$ in the product eigenbasis of $H_0$, we 
use either random Gaussian unitary (GUE) or orthogonal (GOE) matrices. Their 
statistical properties are completely characterized by the following second 
moments (for notational ease we ignore the subscript $_{1,{\rm e}}$ for a
moment)
\begin{equation}
\< V_{ij} V_{kl}\> = \delta_{il}\, \delta_{jk}
   + \chi_{\rm GOE}\; \delta_{ik}\, \delta_{jl} \; ,
\label{B:GEdef}
\end{equation}
where $\chi_{\rm GOE} = 1$ if $V$ is taken from the GOE, while
$\chi_{\rm GOE} = 0$ if $V$ is taken from the GUE. In the interaction picture
we then find:
\begin{equation}
\fl\qquad \< \tilde V_{ij}\, \tilde V_{kl}\> = \big (\, \delta_{il}\,
   \delta_{jk}\; \rme^{-\mimath (E_j-E_i)\, (\tau-\tau')} + \chi_{\rm GOE}\;
   \delta_{ik}\, \delta_{jl}\; \rme^{-\mimath (E_j-E_i)\, (\tau+\tau')}\, \big )
   \; .
   \label{B:tVextdef}
\end{equation}

\paragraph{\boldmath $p[\mathcal{T} \< \tilde V \tilde V'\> \varrho_0\otimes\varrho_0](t)$:}
We first compute the average 
\begin{eqnarray}
\mathcal{T}\< \tilde V\, \tilde V'\, \varrho_0 \otimes \varrho_0\> =
   \sum_{ij\, kl\, mn} |ij\>\; \mathcal{T}\< \tilde V(\tau)_{ij,kl}\;
   \tilde V(\tau')_{kl,mn}\> \; \< mn|\; \varrho_0\otimes\varrho_0 \nonumber\\
\qquad = 
   \sum_{ij\, kl} |ij\> \; \rme^{-\mimath (E_{kl}-E_{ij})\, |\tau-\tau'|}\; 
   \< ij|\; \varrho_0 \otimes \varrho_0 \; ,
\end{eqnarray}
where we have used the averaging rule, \eref{B:tVextdef}, as well as the
fact that the time-ordering operator $\mathcal{T}$ requests to exchange $\tau$
with $\tau'$ whenever $\tau < \tau'$.
The indices $i,k$ and $m$ denote basis states in $\mcH_1$, while the indices 
$j,l$ and $n$ denote basis states in $\mcH_{\rm e}$. We can rewrite that
expression in a more compact form by employing the 
diagonal matrices $\bm{C}_x(\tau)$, defined in \eref{B:CCxdef}:
\begin{equation}
\fl \mathcal{T}\< \tilde V\, \tilde V'\, \varrho_0 \otimes \varrho_0\> =
  \big (\, \bm{C}_1(|\tau-\tau'|) \otimes \bm{C}_{\rm e}(|\tau-\tau'|)\, \big )
   \big (\, \rho_1 \otimes |\psi_{\rm e}\>\, \<\psi_{\rm e}|\, \big )
   \otimes
   \big (\, \rho_1 \otimes |\psi_{\rm e}\>\, \<\psi_{\rm e}|\, \big ).
\end{equation}
Now we apply the partial traces over subsystem $\mcH_1$
\begin{equation}
\tr_1\big (\, \< \tilde V\, \tilde V'\>\, \varrho_0\, \big )\,
\tr_1\, \varrho_0 = C_1(|\tau-\tau'|)\; \bm{C}_{\rm e}(|\tau-\tau'|)\,
   |\psi_{\rm e}\>\, \<\psi_{\rm e}| \; .
\end{equation}
The final trace over the environment yields
\begin{equation}
p[\mathcal{T}\< \tilde V\, \tilde V'\> \varrho_0\otimes\varrho_0](t)=
   C_1(|\tau-\tau'|)\; C_{\rm e}(|\tau-\tau'|) \; .
\end{equation}

\paragraph{\boldmath $p[\< \tilde V' \varrho_0 \tilde V\> \otimes \varrho_0](t)$:} 
We first compute the average 
\begin{eqnarray}
\fl\qquad \<\tilde V'\, \varrho_0\, \tilde V\> \otimes \varrho_0 =
   \sum_{ij\, kl\, mn\, pq} |ij\>\; \left\< \tilde V'_{ij,kl}\; 
   \< kl|\varrho_0|mn\>\; \tilde V_{mn,pq}\right\> \; \< pq| \otimes \varrho_0
\nonumber\\
\fl\qquad\qquad = \sum_{ij\, kl} |ij\>\;
   \rme^{-\mimath (E_{ij}-E_{kl})\, (\tau-\tau')}\;
   \< kl|\varrho_0| kl\>\; \< ij| \otimes \varrho_0 \; ,
\end{eqnarray}
where we have used \eref{B:tVextdef}.
We apply the partial traces over subsystem $\mcH_1$
\begin{eqnarray}
\fl\qquad \tr_1\big (\, \<\tilde V'\, \varrho_0\, \tilde V\>\, \big )\;
\tr_1\, \varrho_0 = \left( {\textstyle\sum_{ik}}
   \rme^{-\mimath (E_i-E_k)\, (\tau-\tau')}\; \< k|\rho_1|k\> \right)
\nonumber\\
\fl\qquad\qquad\qquad\times
   \sum_{jl} |j\>\; \rme^{-\mimath (E_j-E_l)\, (\tau-\tau')}\; 
   \< l|\psi_{\rm e}\>\, \< \psi_{\rm e}|l \>\, \< j|\psi_{\rm e}\>\, 
   \< \psi_{\rm e}| \nonumber\\
\fl\qquad\qquad = C_1(\tau-\tau')\; 
   \sum_{jl} |j\>\; \rme^{-\mimath (E_j-E_l)\, (\tau-\tau')}\;
   \< l|\psi_{\rm e}\>\, \< \psi_{\rm e}|l \>\, \< j|\psi_{\rm e}\>\,
   \< \psi_{\rm e}| \; .
\end{eqnarray}
The final trace over the environment yields
\begin{eqnarray}
\fl p[\<\tilde V' \varrho_0 \tilde V\> \otimes \varrho_0]= C_1(\tau-\tau')\;
   \sum_{jl} \<\psi_{\rm e}|l\> \;\rme^{\mimath E_l\, (\tau-\tau')}\; 
   \< l|\psi_{\rm e}\> \; \<\psi_{\rm e}|j\> \;
   \rme^{-\mimath E_j\, (\tau-\tau')}\; \< j|\psi_{\rm e}\> \nonumber\\
\fl\qquad = C_1(\tau-\tau')\; S_{\rm e}(\tau-\tau') \; ,
\end{eqnarray}
where $S_x(\tau)$ is defined in \eref{B:Sxdef}.

\paragraph{\boldmath $p[\< \tilde V \varrho_0\otimes \tilde V' \varrho_0\>](t)$:} 
We first compute the average
\begin{eqnarray}
\fl\qquad
   \<\tilde V\, \varrho_0\otimes \tilde V'\, \varrho_0\> =
   \sum_{ij\, kl\, mn\, pq} |ij\>\; \left\< \tilde V_{ij,kl}\; \< kl|\,
   \varrho_0 \otimes |mn\>\; \tilde V_{mn,pq}\right\>  \< pq|\,
   \varrho_0 \nonumber\\
\fl\qquad\qquad
= \sum_{ij\, kl} |ij\>\; \rme^{-\mimath (E_{kl}-E_{ij})\, (\tau-\tau')}\;
   \< kl|\; \varrho_0 \otimes |kl\>\, \< ij|\; \varrho_0 \; .
\end{eqnarray}
Then we apply the partial traces over subsystem $\mcH_1$
\begin{eqnarray}
\fl\qquad \big \<\, \tr_1\big (\, \tilde V\, \varrho_0\, \big )\,
   \tr_1\big (\, \tilde V'\, \varrho_0\, \big )\, \big \> = 
\big (\, {\textstyle\sum_{ik}} \rme^{-\mimath (E_k-E_i)\, (\tau-\tau')}\,
   \< k|\rho_1|i\>\, \< i|\rho_1|k\> \, \big ) \nonumber\\
\fl\qquad\qquad\times \sum_{jl} |j\>\; \rme^{-\mimath (E_l-E_j)\, (\tau-\tau')}\;
   \< l|\psi_{\rm e}\> \, \<\psi_{\rm e}|l\> \, \< j|\psi_{\rm e}\> \, 
   \<\psi_{\rm e}| \; .
\end{eqnarray}
The final trace over the environment yields
\begin{equation}
p[\< \tilde V\varrho_0\otimes\tilde V'\varrho_0\>]=
   S_1(\tau-\tau')\; S_{\rm e}(\tau-\tau') \; .
\end{equation}

\paragraph{\boldmath $p[\< \tilde V'\varrho_0\otimes\varrho_0\tilde V\>](t)$:} 
We first compute the average 
\begin{eqnarray}
\fl\qquad \< \tilde V'\, \varrho_0\otimes\varrho_0\, \tilde V\> =
   \sum_{ij\, kl\, mn\, pq} |ij\>\; \left\< \tilde V'_{ij,kl}\; \< kl|\,
   \varrho_0 \otimes \varrho_0\, |mn\>\; \tilde V_{mn,pq}\right\> \< pq|
\nonumber\\
\fl\qquad\qquad = \sum_{ij\, kl} |ij\>\; 
   \rme^{-\mimath (E_{ij}-E_{kl})\, (\tau-\tau')}\; \< kl|\, \varrho_0 \otimes 
   \varrho_0\, |kl\>\; \< ij|
\end{eqnarray}
We apply the partial traces over subsystem $\mcH_1$
\begin{eqnarray}
\fl\qquad \big \<\, \tr_1\big (\, \tilde V\, \varrho_0\, \big )\,   
    \tr_1\big (\, \varrho_0\, \tilde V'\, \big )\, \big \> =
   \big (\, {\textstyle\sum_{ik}} \rme^{-\mimath (E_i-E_k)\, (\tau-\tau')}\,
   \< k|\rho_1|i\>\, \< i|\rho_1|k\> \, \big ) \nonumber\\
\fl\qquad\qquad\times \sum_{jl} |j\>\; \rme^{-\mimath (E_j-E_l)\, (\tau-\tau')}\;
   \< l|\psi_{\rm e}\> \, \<\psi_{\rm e}|l\> \, \< j| \; .
\end{eqnarray}
The final trace over the environment yields
\begin{equation}
p[\< \tilde V\varrho_0\otimes\varrho_0 \tilde V'\>]=
   S_1(\tau-\tau')\; C_{\rm e}(\tau-\tau') \; .
\end{equation}

\subsection{\label{sec:appGOE}  The GOE case}

In the GOE case, the average over the coupling yields, besides the previously
considered GUE-term an additional one; see \eref{B:tVextdef}. In the 
following we will redo the calculation of the previous subsection, but 
consider only that additional term. As a reminder, we add the subscript $_2$
to the brackets which denote the ensemble average.

\paragraph{\boldmath $p[\mathcal{T}\< \tilde V \tilde V'\>_2 \varrho_0\otimes\varrho_0](t)$:}
We first compute the average
\begin{eqnarray}
\fl\qquad
  \mathcal{T}\< \tilde V\, \tilde V'\>_2 \, \varrho_0 \otimes \varrho_0 =
  \sum_{ij\, kl\, mn} |ij\>\; \mathcal{T}\< \tilde V(\tau)_{ij,kl}\;
  \tilde V(\tau')_{kl,mn}\>_2\; \< mn|\; \varrho_0\otimes\varrho_0\nonumber\\
\fl\qquad\qquad =
   \sum_{ij} |ij\> \; \< ij|\; \varrho_0 \otimes \varrho_0 
 = \varrho_0 \otimes \varrho_0\; ,
\end{eqnarray}
where we have used the averaging rule, \eref{B:tVextdef}. Here the 
time-ordering operator has no effect.
Now we apply the partial traces over subsystem $\mcH_1$
\begin{equation}
\tr_1\big (\, \< \tilde V\, \tilde V'\>_2 \, \varrho_0\, \big )\,
\tr_1\, \varrho_0 = |\psi_{\rm e}\>\, \<\psi_{\rm e}| \; .
\end{equation}
The final trace over the environment yields
\begin{equation}
p[\< \tilde V\, \tilde V'\>_2 \varrho_0\otimes\varrho_0](t)= 1 \; .
\end{equation}

\paragraph{\boldmath $p[\< \tilde V' \varrho_0 \tilde V\>_2 \otimes \varrho_0](t)$:} 
We first compute the average
\begin{eqnarray}
\fl\qquad \<\tilde V'\, \varrho_0\, \tilde V\>_2 \otimes \varrho_0 =
   \sum_{ij\, kl\, mn\, pq} |ij\>\; \left\< \tilde V'_{ij,kl}\; 
   \< kl|\varrho_0|mn\>\; \tilde V_{mn,pq}\right\>_2\; \< pq| \otimes\varrho_0
\nonumber\\
\fl\qquad\qquad = \sum_{ij\, kl} |ij\>\;
   \rme^{-\mimath (E_{ij}-E_{kl})\, (\tau+\tau')}\;
   \< kl|\varrho_0| ij\>\; \< kl| \otimes \varrho_0 \; ,
\end{eqnarray}
where we have used \eref{B:tVextdef}.
We apply the partial traces over subsystem $\mcH_1$
\begin{eqnarray}
\fl\qquad \tr_1\big (\, \<\tilde V'\, \varrho_0\, \tilde V\>_2\, \big )\;
\tr_1\, \varrho_0 = 
   \sum_{jl} |j\>\; \rme^{-\mimath (E_j-E_l)\, (\tau+\tau')}\; 
   \< l|\psi_{\rm e}\>\, \< \psi_{\rm e}|j \>\, \< l|\psi_{\rm e}\>\, 
   \< \psi_{\rm e}| \; .
\end{eqnarray}
The final trace over the environment yields
\begin{eqnarray}
\fl p[\<\tilde V' \varrho_0 \tilde V\> \otimes \varrho_0]=
   \sum_{jl} \< l|\psi_{\rm e}\> \;\rme^{\mimath E_l\, (\tau+\tau')}\; 
   \< l|\psi_{\rm e}\> \; \<\psi_{\rm e}|j\> \;
   \rme^{-\mimath E_j\, (\tau+\tau')}\; \<\psi_{\rm e}|j\> \nonumber\\
\fl\qquad = S'_{\rm e}(-\tau-\tau') \; ,
\end{eqnarray}
where $S'_x(\tau)$ is defined in \eref{B:Spxdef}.

\paragraph{\boldmath $p[\< \tilde V \varrho_0\otimes \tilde V' \varrho_0\>_2](t)$:} 
We first compute the average 
\begin{eqnarray}
\fl\qquad
   \<\tilde V\, \varrho_0\otimes \tilde V'\, \varrho_0\>_2 =
   \sum_{ij\, kl\, mn\, pq} |ij\>\; \left\< \tilde V_{ij,kl}\; \< kl|\,
   \varrho_0 \otimes |mn\>\; \tilde V_{mn,pq}\right\>_2  \< pq|\,
   \varrho_0 \nonumber\\
\fl\qquad\qquad
= \sum_{ij\, kl} |ij\>\; \rme^{-\mimath (E_{kl}-E_{ij})\, (\tau+\tau')}\;
   \< kl|\; \varrho_0 \otimes |ij\>\, \< kl|\; \varrho_0 \; .
\end{eqnarray}
Then we apply the partial traces over subsystem $\mcH_1$
\begin{eqnarray}
\fl\qquad \big \<\, \tr_1\big (\, \tilde V\, \varrho_0\, \big )\,
   \tr_1\big (\, \tilde V'\, \varrho_0\, \big )\, \big \>_2 = 
\big (\, {\textstyle\sum_{ik}} \rme^{-\mimath (E_k-E_i)\, (\tau+\tau')}\,
   \< k|\rho_1|i\>\, \< k|\rho_1|i\> \, \big ) \nonumber\\
\fl\qquad\qquad\times \sum_{jl} |j\>\; \rme^{-\mimath (E_l-E_j)\, (\tau+\tau')}\;
   \< l|\psi_{\rm e}\> \, \<\psi_{\rm e}|j\> \, \< l|\psi_{\rm e}\> \, 
   \<\psi_{\rm e}| \; .
\end{eqnarray}
The final trace over the environment yields
\begin{equation}
p[\< \tilde V\varrho_0\otimes\tilde V'\varrho_0\>_2]=
   S'_1(\tau+\tau')\; S'_{\rm e}(\tau+\tau') \; .
\end{equation}

\paragraph{\boldmath $p[\< \tilde V'\varrho_0\otimes\varrho_0\tilde V\>_2](t)$:} 
We first compute the average 
\begin{eqnarray}
\fl\qquad \< \tilde V'\, \varrho_0\otimes\varrho_0\, \tilde V\>_2 =
   \sum_{ij\, kl\, mn\, pq} |ij\>\; \left\< \tilde V'_{ij,kl}\; \< kl|\,
   \varrho_0 \otimes \varrho_0\, |mn\>\; \tilde V_{mn,pq}\right\>_2 \< pq|
\nonumber\\
\fl\qquad\qquad = \sum_{ij\, kl} |ij\>\; 
   \rme^{-\mimath (E_{ij}-E_{kl})\, (\tau+\tau')}\; \< kl|\, \varrho_0 \otimes 
   \varrho_0\, |ij\>\; \< kl| \; .
\end{eqnarray}
We apply the partial traces over subsystem $\mcH_1$
\begin{eqnarray}
\fl\qquad \big \<\, \tr_1\big (\, \tilde V\, \varrho_0\, \big )\,   
    \tr_1\big (\, \varrho_0\, \tilde V'\, \big )\, \big \>_2 =
   \big (\, {\textstyle\sum_{ik}} \rme^{-\mimath (E_i-E_k)\, (\tau+\tau')}\,
   \< k|\rho_1|i\>\, \< k|\rho_1|i\> \, \big ) \nonumber\\
\fl\qquad\qquad\times \sum_{jl} |j\>\; \rme^{-\mimath (E_j-E_l)\, (\tau+\tau')}\;
   \< l|\psi_{\rm e}\> \, \<\psi_{\rm e}|j\> \, \< l| \; .
\end{eqnarray}
The final trace over the environment yields
\begin{equation}
p[\< \tilde V\varrho_0\otimes\varrho_0 \tilde V'\>_2]=
   S'_1(-\tau-\tau')\; .
\end{equation}

\section[\\Some particular correlation functions]{\label{aB} Some particular correlation functions relevant for the average purity decay}
\subsection{Definitions}

Averaging over the perturbation (\ie the coupling) leads to expressions 
which may involve the diagonal matrix
\begin{equation}
  \bm{C}_x(\tau)= \sum_{ik} |i\>\; \rme^{-\mimath (E_k-E_i)\, \tau}\; \< i|
  \qquad x\in 1,{\rm e}\; .
  \label{B:CCxdef}
\end{equation}
Here, $x$ denotes one of the two subsystems considered in \ref{sec:aA}, either 
the qubit or the environment. Evidently,
the energies $E_k$ are the eigenvalues of the corresponding Hamiltonian. In
the derivations in~\ref{sec:aA}, the expectation value of $\bm{C}_x(\tau)$ with
respect to the initial state $\varrho_0=\rho_1\otimes\rho_\e$ are of 
particular importance. These are denoted by
\begin{equation}\label{B:Cxdef}
  C_x(\tau)= \tr\big (\, \bm{C}_x(\tau)\; \rho_x\, \big )\qquad x\in 1,{\rm e} \; .
\end{equation}
In this work, $\rho_{\rm e}= |\psi_\e\> \<\psi_\e|$ is always a pure state.

We will also encounter another type of correlation function, which may be
defined as follows
\begin{equation}\label{B:Sxdef}
  S_x(\tau)= \tr_x\big [\, \rme^{-\mimath H_x\, \tau}\, \rho_x\, 
     \rme^{\mimath H_x\, \tau}\, \rho_x\, \big ] \; .
\end{equation}
If the initial state $\rho_x$ is pure, this quantity becomes the return probability.
At $\tau=0$ it gives the
purity of $\rho_x$. In the case of GOE averages, we also 
encounter the correlation function
\begin{equation}\label{B:Spxdef}
  S'_x(\tau)= \tr_x\big [\, \rme^{-\mimath H_x\, \tau}\, \rho_x\,
     \rme^{\mimath H_x\, \tau}\, \rho_x^T, \big ] \; .
\end{equation}

\subsection{Properties}
\paragraph{The limit of infinite dimension (of $\mcH_{\rm e}$):}
In the main part of this paper, we focus on the limit, where the dimension of 
the environment(s) becomes infinite. In that case it makes sense to perform
an additional spectral average over $H_\e$ and/or $H_{\e'}$. In the case of 
$\bm{C}_\e(t)$ this yields
\begin{equation}\label{B:Cbardef}
\fl\qquad  \< \bm{C}_{\rm e}(\tau)\> = \bar C(\tau)\; \openone_{\rm e} \qquad
  \bar C(\tau)= \< C_\e(\tau)\> = 1+ \delta(\tau/\tau_H) 
      - b_2^{(\beta)}(\tau/\tau_H) \; ,
\end{equation}
where $b_2^{(\beta)}(t)$ is the two-point spectral form factor with time
measured in units of the Heisenberg time $\tau_H$ for the corresponding 
spectral ensemble. In the limit of large dimension $N_\e=\dim(\mcH_\e)$ we find 
for the other correlation functions:
\begin{eqnarray}
\fl\qquad S_\e(\tau)= \tr\left( 
   \rme^{-\mimath H_\e \tau} |\psi_\e\> \<\psi_\e|
   \rme^{\mimath H_\e\tau} |\psi_\e\> \<\psi_\e| \right) \\
\fl\qquad S'_\e(\tau)= \tr\left(
   \rme^{-\mimath H_\e \tau} |\psi_\e\> \<\psi_\e|
   \rme^{\mimath H_\e\tau} |\psi_\e^*\> \<\psi_\e^*| \right) \; .
\end{eqnarray}
For random pure states, as considered in the present work, both correlation
functions are at most of order one at $\tau=0$. For $\tau>0$ they drop very
quickly and soon become of order
$N_\e^{-1}$. This happens on the same time scale,
where $C_\e(\tau)$ drops from values of the order $N_\e$ to values of the order
one. In that sense we consider these correlation functions to contribute only 
$\Or(N_\e^{-1})$ corrections to the result given in 
\eref{B:pulrires}.

\paragraph{The different correlation functions for a single qubit}
A single qubit is 
a two level system. The most general pure initial state is given by 
$|a\>= |1\>\, a_1 + |2\>\, a_2$, with $|a_1|^2 + |a_2|^2 = 1$. The most general
mixed state is given by 
$\rho_1= \lambda_1\, |a\>\, \<a| + \lambda_2\, |b\>\, \<b|$, with 
$\lambda_1, \lambda_2 \ge 0$, real, $\lambda_1+\lambda_2=1$, and 
$|a\>,\, |b\>$ arbitrary pure states with $\< a|b\> =0$. We will now 
investigate the behaviour of the different correlation functions 
${\rm Re} C_1(\tau), S_1(\tau)$, and $S_1'(\tau)$ as it depends on the 
initial state and the Hamiltonian $H_1= |1\> E_1 \< 1| + |2\> E_2 \< 2|$.

\begin{itemize}
\item[(a)] Assume $\rho_1= |a\>\, \< a|$. Then
\begin{equation}
\fl\quad
C_1(\tau)= \sum_{ik} |a_i|^2\, \rme^{-\mimath (E_k-E_i)\, \tau}
 = 1 + \cos\Delta\tau +\mimath (|a_2|^2 - |a_1|^2)\, \sin\Delta\tau \; ,
\end{equation}
holds, so
\begin{equation}
\fl\quad {\rm Re}\, C_1(\tau)= 1 + \cos\Delta\tau \qquad \Delta= E_2-E_1 \; .
\end{equation}

\item[(b)] For $\rho_1= \lambda_1\, |a\>\, \<a| + \lambda_2\, |b\>\, \<b|$ we
still have
\begin{equation}\label{aB:ReC1}
  \fl\quad {\rm Re}\, C_1(\tau)= 1 + \cos\Delta\tau \; .
\end{equation}
\item[(c)] Assume $\rho_1= |a\>\, \< a|$. Then we find for 
$S_1(\tau)= s(|a\>; \tau)$
\begin{equation}
\fl\quad
S_1(\tau)= \sum_{ik} |a_i|^2 |a_k|^2\, \rme^{-\mimath (E_i-E_k)\, \tau}
 = |a_1|^4 + |a_2|^4 + 2\, |a_1|^2 |a_2|^2\, \cos\Delta\tau \; .
\end{equation}
This expression only depends on the absolute values squared of the coefficients
$a_1$ and $a_2$. Therefore we may parametrize them without loss of generality 
as $a_1= \cos\phi$ and $a_2= \sin\phi$. We then find
\begin{equation}\label{aB:S1pure}
\fl\quad
S_1(\tau)= g_\phi + (1- g_\phi)\, \cos\Delta\tau\qquad
g_\phi= |a_1|^4 + |a_2|^4= 1- \frac{1}{2}\; \sin^2\, 2\phi \; .
\end{equation}

\item[(d)] For a general mixed state 
$\rho_1= \lambda_1\, |a\>\, \<a| + \lambda_2\, |b\>\, \<b|$ we find
\begin{eqnarray}
\fl\quad
S_1(\tau)= \lambda_1^2\, s(|a\>; \tau) + \lambda_2^2\, s(|b\>; \tau) + 
   2\, \lambda_1\lambda_2\, \sum_{ik} a_i a_k^* b_k b_i^*\, \cos (E_i-E_k)\tau
\nonumber\\
\fl\quad\qquad = \lambda_1^2\, s(|a\>; \tau) + \lambda_2^2\, s(|b\>; \tau) +
   4\, \lambda_1\lambda_2\, |a_1|^2 |b_1|^2 \big (\, 1 - \cos\Delta\tau\, \big )
   \; ,
\end{eqnarray}
where we have used that $\< a|b\> = a_1 b_1^* + a_2 b_2^* = 0$.
Note that the coefficients $a_1, a_2, b_1, b_2$ may be arranged into a square
unitary matrix, and must therefore be of the following general form
\begin{equation}\label{aB:abunit}
\left(\!\!\!\begin{array}{cc} a_1 & b_1\\ a_2 & b_2 \end{array}\!\!\!\right)=
\rme^{\mimath\vartheta}\left(\!\!\!\begin{array}{cc} 
   \rme^{\mimath\xi}\, \cos\phi & \rme^{\mimath\chi}\, \sin\phi\\
   -\, \rme^{-\mimath\chi}\, \sin\phi & \rme^{-\mimath\xi}\, \cos\phi 
\end{array}\!\!\!\right) \; .
\end{equation}
This shows that 
$s(|a\>; \tau)= s(|b\>; \tau) = g_\phi + (1- g_\phi)\, \cos\Delta\tau$, and that
\begin{equation}
\fl\quad
S_1(\tau)= (\lambda_1^2 + \lambda_2^2) \big [\, 
   g_\phi + (1-g_\phi)\, \cos\Delta\tau\, \big ] + 2\, \lambda_1\lambda_2\, 
   (1-g_\phi)\, \big (\, 1 - \cos\Delta\tau\, \big ) \; .
\end{equation}
Since $\lambda_1+\lambda_2= 1$ it is convenient to set $\lambda_1= \cos^2\theta$
and $\lambda_2= \sin^2\theta$ such that we may write
$\lambda_1^2 + \lambda_2^2 =g_\theta$ and obtain
\begin{equation}\label{aB:S1mixed}
\fl\quad
S_1(\tau)= 1- g_\theta- g_\phi + 2\, g_\theta g_\phi + (2g_\theta-1) (1-g_\phi)\, 
   \cos\Delta\tau \; .
\end{equation}

\item[(e)] Assume again that $\rho_1= |a\>\, \<a|$. Then we find for 
$S'_1(\tau)= s'(|a\>; \tau)$
\begin{eqnarray}
\fl\quad
S'_1(\tau)= |a_1|^4 + |a_2|^4 + a_1^2 (a_2^*)^2\, \rme^{\mimath\Delta\tau} + 
   a_2^2 (a_1^*)^2\, \rme^{-\mimath\Delta\tau} \nonumber\\
\fl\quad\qquad
 = g_\phi + 2\, |a_1|^2 |a_2|^2\, \cos(\Delta\tau+2\eta)\nonumber\\
\fl\quad\qquad
 = g_\phi + (1-g_\phi)\, \cos(\Delta\tau+2\eta)\qquad
\eta= {\rm arg}(a_1) - {\rm arg}(a_2) \; .
\label{aB:S1ppure}\end{eqnarray}
However, as discussed in \sref{sec:goeone}, a natural symmetry around
the $y$ axis (in the Bloch sphere picture) appears for $\Delta=0$. In that case, using $\gamma$ as defined
in that section, one can prove that
\begin{equation}
	1-S'_1(\tau)\big|_{\Delta=0}=(1-g_\phi)[1- \cos(2\eta)]=\sin^2\gamma
	\label{aB:S1ppurenD}
\end{equation}
using elementary geometric and trigonometric considerations. 

\item[(f)] For a general mixed state
$\rho_1= \lambda_1\, |a\>\, \<a| + \lambda_2\, |b\>\, \<b|$ we find
\begin{equation}
\fl\quad
S'_1(\tau)= \lambda_1^2\, s'(|a\>; \tau) + \lambda_2^2\, s'(|b\>; \tau)
 + 2\, \lambda_1\lambda_2\, {\rm Re}\, \sum_{ik} a_i a_k^* b_k^* b_i\, 
   \rme^{-\mimath (E_i-E_k)\, \tau} \; .
\end{equation}
It follows from \eref{aB:abunit} that 
${\rm arg}(b_1) - {\rm arg}(b_2) = {\rm arg}(a_1) - {\rm arg}(a_2) - \pi$, such
that the equality $s'(|a\>; \tau)= s'(|b\>; \tau)$ holds, just as in the case of
$S_1(\tau)$. Therefore we may write
\begin{eqnarray}
\fl\quad
S'_1(\tau)= g_\theta \big [\, g_\phi + (1-g_\phi)\, \cos(\Delta\tau +2\eta)\, 
   \big ] + 2\, \lambda_1\lambda_2\, (1-g_\phi)\, {\rm Re}\big (\, 1 + 
   \rme^{\mimath (\Delta\tau +2\eta -\pi)}\, \big ) \nonumber\\
\fl\quad\qquad
 = 1- g_\theta- g_\phi + 2\, g_\theta g_\phi + (2 g_\theta -1)(1-g_\phi)\,
   \cos(\Delta\tau +2\eta) \; .
\label{aB:S1pmixed}\end{eqnarray}
Note that the only difference to $S(\tau)$ in case (d) is the additional phase
$2\eta$ in the argument of the cosine function. Using the same angle $\gamma$ 
defined with \eref{eq:psi12goe}, in complete analogy with \eref{eq:initialOneGOE},
and using \eref{aB:S1ppurenD} we obtain
\begin{equation}
	1-S'_1(\tau)\big|_{\Delta=0}=1-g_\theta+(2g_\theta-1)\sin^2\gamma.
	\label{aB:S1pmixednD}
\end{equation}

\end{itemize}

\section[\\The exponentiation]{The exponentiation}\label{sec:exponentiation}

The linear response formulae derived throughout the article are expected to work
for high purities or, equivalently, when the first two terms in the
series \eref{eq:bornexpansion} approximate the echo operator well;
it is of obvious interest to extend them to
cover a larger range. The extension of fidelity linear response formulae has,
in some cases, been done with some effort using super-symmetry techniques. This has been
possible partly due to the simple structure of fidelity, but trying to use this
approach for a more complicated object such as purity seems to be out of reach for
the time being.  Exponentiating the formulae obtained from the linear
response formalism has proven to be in good agreement with the exact
super-symmetric and/or numerical results for  fidelity, if the perturbation 
is not too big. The exponentiation of the
linear  response  formulae  for  purity  can  be  compared  with  Monte  Carlo
simulations in order to prove its  validity.  We wish to explain the details
required to implement this procedure in this appendix. 

Given a linear response formula $P_{\rm LR}(t)$ (for which $P(0)=1$), and an
expected asymptotic value for infinite time $P_\infty$, the exponentiation
reads
\begin{equation}
  \label{eq:ELRextension}
  P_{\rm ELR}(t)=P_\infty+(1-P_\infty)\exp\left[- \frac{1-P_{\rm LR}(t)}{1-P_\infty}\right].
\end{equation}
This particular form guaranties that $P_{\rm ELR}$ equals $P_{\rm LR}$ for
short times, and that $\lim_{t \to \infty}P_{\rm ELR}(t)=P_\infty$ The
particular value of $P_\infty$ will depend on the physical situation; in our
case it will depend on the configuration and on the initial conditions.

Let us write the initial condition as
\begin{equation}
 |\psi_{12}(\theta)\>=\cos\theta|\tilde 0_1 \tilde0_2\>+\sin\theta|\tilde 1_1 \tilde 1_2\>
\end{equation}
for some rotated qubits $|\tilde 0_i\>$, $|\tilde 1_i\>$.  In each of the
qubits interacting with the environment we assumed that for long enough times,
the totally depolarizing channel $\mathcal E_{\rm d}$ will act (recall that the
totally depolarizing channel is defined as $\mathcal E_{\rm
d}[\rho]=\openone/\tr \openone$ for any density matrix $\rho$). 
Hence, for the spectator configuration, the final value of purity is
\begin{equation}
P_\infty=
 P\left( \mathcal E_{\rm d}\otimes \openone [|\psi_{12}(\theta)\>\<\psi_{12}(\theta)|] \right)
=\frac{g_\theta}{2}
\end{equation}
whereas for both the joint  and the separate environment configuration we use
the value 
\begin{equation}
P_\infty
   =P\left(\mathcal E_{\rm d}\otimes\mathcal E_{\rm d} [|\psi_{12}(\theta)\>\<\psi_{12}(\theta)|]\right) =\frac{1}{4}.
\end{equation}
Good   agreement   is  found   with   Monte Carlo   simulations  for   moderate
and strong ºcouplings.

\section*{References}
\bibliographystyle{unsrt}
\bibliography{paperdef,miblibliografia,specialbib}

\begin{thebibliography}{10}

\bibitem{NatureTomography}
H.~{H{\"a}ffner}, W.~{H{\"a}nsel}, C.~F. {Roos}, J.~{Benhelm},
  D.~{Chek-al-kar}, M.~{Chwalla}, T.~{K{\"o}rber}, U.~D. {Rapol}, M.~{Riebe},
  P.~O. {Schmidt}, C.~{Becher}, O.~{G{\"u}hne}, W.~{D{\"u}r}, and R.~{Blatt}.
\newblock {Scalable multiparticle entanglement of trapped ions}.
\newblock {\em Nature}, 438:643, December 2005.

\bibitem{fourparticleentanglement}
C.~A. Sackett, D.~Kielpinski, B.~E. King, C.~Langer, V.~Meyer, C.~J. Myatt,
  M.~Rowe, Q.~A. Turchette, W.~M. Itano, D.~J. Wineland, and C.~Monroe.
\newblock Experimental entanglement of four particles.
\newblock {\em Nature}, 404:256, Mar 2000.

\bibitem{photonentanglement}
Zhi Zhao, Yu-Ao Chen, An-Ning Zhang, Tao Yang, Hans~J. Briegel, and Jian-Wei
  Pan.
\newblock Experimental demonstration of five-photon entanglement and
  open-destination teleportation.
\newblock {\em Nature}, 430:54, Jul 2004.

\bibitem{antidecoherenceentanglement}
H.~H{\"a}ffner, F.~Schmidt-Kaler, W.~H{\"a}nsel, C.~F. Roos, T.~K{\"o}rber,
  M.~Chwalla, M.~Riebe, J.~Benhelm, U.~D. Rapol, C.~Becher, and R.~Blatt.
\newblock Robust entanglement.
\newblock {\em Applied Physics B: Lasers and Optics}, 81:151, Jul 2005.

\bibitem{1367-2630-6-1-020}
T.~Gorin, T.~Prosen, and T.~H. Seligman.
\newblock A random matrix formulation of fidelity decay.
\newblock {\em New J. Phys.}, 6:20, 2004.

\bibitem{reflosch}
Thomas Gorin, Toma{\v z} Prosen, Thomas~H. Seligman, and Marko {\v Z}nidari{\v
  c}.
\newblock {\em Phys. Rep.}, 435:33, 2006.

\bibitem{expRudi}
R.~Sch\"{a}fer, H.-J. St\"{o}ckmann, T.~Gorin, and T.~H. Seligman.
\newblock Experimental verification of fidelity decay: From perturbative to
  {F}ermi golden rule regime.
\newblock {\em Phys. Rev. Lett.}, 95(18):184102, 2005.

\bibitem{1367-2630-7-1-152}
R.~Sch\"{a}fer, T.~Gorin, T.~H. Seligman, and H.-J. St\"{o}ckmann.
\newblock Fidelity amplitude of the scattering matrix in microwave cavities.
\newblock {\em New J. Phys.}, 7:152, 2005.

\bibitem{gorin-weaver}
T.~Gorin, T.~H. Seligman, and R.~L. Weaver.
\newblock Scattering fidelity in elastodynamics.
\newblock {\em Phys. Rev. E}, 73(1):015202(R), 2006.

\bibitem{Zur91}
W.H. Zurek.
\newblock Decoherence and the transition from quantum to classical.
\newblock {\em Phys. Today}, 44(10):36, 1991.

\bibitem{gpss2004}
T.~Gorin, T.~Prosen, T.~H. Seligman, and W.~T. Strunz.
\newblock Connection between decoherence and fidelity decay in echo dynamics.
\newblock {\em Phys. Rev. A}, 70(4):042105, 2004.

\bibitem{1464-4266-4-4-325}
T.~Gorin and T.~H. Seligman.
\newblock A random matrix approach to decoherence.
\newblock {\em J. Opt. B}, 4(4):S386, 2002.

\bibitem{pinedaRMTshort}
Carlos Pineda and Thomas~H. Seligman.
\newblock Bell pair in a generic random matrix environment.
\newblock {\em Phys. Rev. A}, 75(1):012106, 2007.

\bibitem{privatepineda2006}
C. Pineda, private communication, 2006.

\bibitem{zurekreview}
Wojciech~Hubert Zurek.
\newblock Decoherence, einselection, and the quantum origins of the classical.
\newblock {\em Rev. Mod. Phys.}, 75(3):715, May 2003.

\bibitem{purityfidelity}
Toma{\v z} Prosen and Thomas~H Seligman.
\newblock Decoherence of spin echoes.
\newblock {\em J. Phys. A}, 35(22):4707, 2002.

\bibitem{shortRMT}
H.-J. {St{\"o}ckmann} and R.~{Sch{\"a}fer}.
\newblock {Fidelity Recovery in Chaotic Systems and the Debye-Waller Factor}.
\newblock {\em Phys. Rev. Lett.}, 94(24):244101, June 2005.

\bibitem{gorin:244105}
T.~Gorin, H.~Kohler, T.~Prosen, T.~H. Seligman, H.-J.~St\" ockmann, and M.~{\v
  Z}nidari{\v c}.
\newblock Anomalous slow fidelity decay for symmetry-breaking perturbations.
\newblock {\em Phys. Rev. Lett.}, 96(24):244105, 2006.

\bibitem{Pineda01}
G.~L. {Celardo}, C.~{Pineda}, and M.~{{\v Z}nidari{\v c}}.
\newblock {Stability of quantum Fourier transformation on Ising quantum
  computer}.
\newblock {\em Int. J. Quantum Inf.}, 3(3):441, 2005.

\bibitem{cartanRMT}
{\` E}lie Cartan.
\newblock Quasi composition algebras.
\newblock {\em Abh. Math. Sem. Hamburg}, 11:116, 1935.

\bibitem{mehta}
Madan~Lal Mehta.
\newblock {\em Random Matrices}.
\newblock Academic Press, San Diego, California, second edition, 1991.

\bibitem{wei:022110}
Tzu-Chieh Wei, Kae Nemoto, Paul~M. Goldbart, Paul~G. Kwiat, William~J. Munro,
  and Frank Verstraete.
\newblock Maximal entanglement versus entropy for mixed quantum states.
\newblock {\em Phys. Rev. A}, 67(2):022110, 2003.

\bibitem{andrereview}
Florian Mintert, Andr\' e~R~R~Carvalho, Marek~K\' us, and Andreas Buchleitner.
\newblock Measures and dynamics of entangled states.
\newblock {\em Phys. Rep.}, 415(4):207, 2005.

\bibitem{firstconcurrence}
Scott Hill and William~K. Wootters.
\newblock Entanglement of a pair of quantum bits.
\newblock {\em Phys. Rev. Lett.}, 78(26):5022, 1997.

\bibitem{wootters}
William~K. Wootters.
\newblock Entanglement of formation of an arbitrary state of two qubits.
\newblock {\em Phys. Rev. Lett.}, 80(10):2245, 1998.

\bibitem{0305-4470-35-6-309}
Toma{\v z} Prosen and Marko {\v Z}nidari{\v c}.
\newblock Stability of quantum motion and correlation decay.
\newblock {\em J. Phys. A}, 35(6):1455, 2002.

\bibitem{pineda:012305}
Carlos Pineda and Thomas~H. Seligman.
\newblock Evolution of pairwise entanglement in a coupled n-body system.
\newblock {\em Phys. Rev. A}, 73(1):012305, 2006.

\bibitem{ziman:052325}
Mario Ziman and Vladimir Buzek.
\newblock Concurrence versus purity: Influence of local channels on bell states
  of two qubits.
\newblock {\em Phys. Rev. A}, 72(5):052325, 2005.

\bibitem{GPS-letter}
T.~{Gorin}, C.~{Pineda}, and T.~H. {Seligman}.
\newblock 2007.
\newblock To be published.

\end{thebibliography}
\end{document}